\documentclass{raa}

\usepackage{graphicx,times}             
\usepackage{natbib}

\begin{document}

   \title{Analysis of a selected sample of RR Lyrae stars in LMC from OGLE III
}

   \volnopage{Vol.0 (200x) No.0, 000--000}      
   \setcounter{page}{1}          

   \author{B.-Q. Chen
      \inst{1,2}
   \and B.-W. Jiang
      \inst{1}
   \and M. Yang
      \inst{1}
   }
\institute{Department of Astronomy, Beijing Normal University, Beijing 100875, P.R.China;{\it bchen@mail.bnu.edu.cn}, {\it bjiang@bnu.edu.cn},
 {\it myang@mail.bnu.edu.cn}
    \\
        \and
Institut Utinam, CNRS UMR6213, OSU THETA, Universit\'e de Franche-Comt\'e, 41bis avenue de l'Observatoire, 25000 Besan\c{c}on, France
   }

   \date{Received~~2009 month day; accepted~~2009~~month day}

\abstract{A systematic study of RR Lyrae stars is performed based on a selected sample
of 655 objects in the Large Magellanic Cloud with observation of long span and numerous
measurements by the Optical Gravitational Lensing
Experiment III project.  The Phase Dispersion Method and linear superposition of the
harmonic oscillations are used to derive the pulsation frequency and variation
properties. It is found that there exists an Oo I and Oo II dichotomy in the LMC RR
Lyrae stars. Due to our strict criteria to identify a frequency, a lower limit
of the incidence rate of Blazhko modulation in LMC is estimated in various
subclasses of RR Lyrae stars.  For fundamental-mode RR Lyrae stars,  the rate
7.5$\%$ is smaller than previous result. In the case of the first-overtone RR
Lyr variables, the rate 9.1$\%$ is relatively high. In addition to the Blazhko
variables, fifteen objects are identified to pulsate in the
fundamental/first-overtone double mode. Furthermore, four objects show a period
ratio around 0.6 which makes them very likely the rare pulsators in the
fundamental/second-overtone double-mode.}


   \authorrunning{B.-Q. Chen, B.-W Jiang \& M.Yang }            
   \titlerunning{RR Lyrae stars in LMC from OGLE III}  

   \maketitle

\section{Introduction}

RR Lyrae stars (RRLS) are pulsating variables on the horizontal branch in the
H-R diagram. They have short periods of 0.2 to 1 day and low metal abundances Z
of 0.00001 to 0.01. Usually they can be easily identified by their light curves and color-color diagrams \citep{2011RAA....11..833L}.
 RRLS is famous for the "Blazhko effect"
\citep{1907AN....175..325B}, a periodic modulation of the amplitude and
phase in the light curves, which is still a mystery in theory today. The main
photometric feature of the Blazhko effect is that the frequency spectra of the
light curves are usually strongly dominated by a symmetric pattern around the
main pulsation frequency $f_{\rm 0}$, i.e., $k f_{\rm 0}$ and $k f_{\rm 0}\pm
f_{\rm BL}$ where $f_{\rm BL}$ is the modulation frequency and $k$ is the
harmonic number \citep{2009AIPC.1170..261K}. Moreover, the higher-order
multiplets such as quintuplets and higher, i.e. multiplets $k f_{\rm 0}\pm l
f_{\rm BL}$, are now also attributed to the Blazhko effect
\citep{2011arXiv1106.4914B}. On the other hand, the amplitudes of modulation
components are usually asymmetric so that one side could be under the detection
limit in highly asymmetric cases, which may lead to the asymmetric appearance of
the frequency spectra. Several models are proposed to explain this effect, including the non-radial resonant rotator/pulsator \citep{1994A&A...291..481G},
the magnetic oblique rotator/pulsator  \citep{2000ASPC..203..299S}, 2:1
resonance model \citep{1980AcA....30..393B},  resonance between radial and
non-radial mode \citep{2004AcA....54..363D}, 9:2 resonance model
\citep{2011ApJ...731...24B} and convective cycles model
\citep{2006ApJ...652..643S}. However, none of them is able to interpret all the
observational phenomena about the Blazhko effect.

A systematic study of RRLS helps to understand their nature, such as the
incidence rate of various pulsation modes, the distribution of modulation
frequency and amplitude and the dependence of the variation properties on
environment. This has been performed on the basis of some datasets with large
amount of data for variables. The early micro-lensing projects, MACHO
\citep{2003ApJ...598..597A} and OGLE \citep{2008AcA....58..163S} that surveyed
mainly LMC, SMC and the Galactic bulge, and the variables-oriented all-sky
survey project, ASAS \citep{2007MNRAS.377.1263S},  have given particularly great
support to such study. Using the MACHO data, \cite{2000ApJ...542..257A} and
\cite{2006A&A...454..257N} analyzed the frequency of 1300 first
overtone RRLS in LMC with an incidence rate of the Blazhko variables of 7.5\%.
Meanwhile, \cite{2003ApJ...598..597A} made a frequency analysis of 6391
fundamental mode RRLS in LMC that resulted in an incidence rate of the Blazhko
variables of 11.9\%.  With the OGLE-I data, \cite{2003A&A...398..213M} searched for multiperiodic pulsators among  38 RRLS in the Galactic Bulge.
\cite{2003AcA....53..307M} made a complete search for multi-period RRLS from
the OGLE-II database, while \cite{2006ApJ...651..197C} presented a catalogue of
1888 fundamental mode RRLS  in the Galactic bulge from the same database.

Recently, \cite{2008CoAst.157..345M} conducted a systematic search for
multiperiod RRLS in $\omega$ Centauri, a globular cluster, and found the
incidence rate of Blazhko modulation pretty high, about 24\% and 38\% for the
fundamental and first-overtone RRLS respectively. \cite{2009MNRAS.400.1006J}
got a 47\% incidence rate in the dedicated Konkoly survey sample of 30
fundamental-mode RRLS in the Galactic field.  \cite{2010ApJ...713L.198K} also
claimed at least 40\% RRLS in the  Kepler space mission sample of 28 objects
exhibiting the modulation phenomenon.

The OGLE-III database released in 2009 \citep{2009AcA....59....1S} contained
24906 light curves that are preliminarily classified as RRLS in the Large
Magellanic Cloud, the ever largest sample of RRLS. This database covers a time
span of about 10 years that makes a large-scale analysis of the RRLS
variation possible. \cite{2009AcA....59....1S} analyzed the basic
statistical features of RRLS in the LMC and divided them into 4 subtypes: RRab,
RRc, RRd and RRe. With this sample of RRLS, the study of the structure of LMC
was developed. \cite{2009ApJ...704.1730P} investigated the structure of the LMC
stellar halo; \cite{2009A&A...503L...9S} found that RRLS in the inner LMC trace
the disk and probably the inner halo; \cite{2010MNRAS.408L..76F} established a
small but significant radial gradient in the mean periods of Large Magellanic
Cloud (LMC) RR Lyrae variables. The data of RRLS in SMC and the Galactic bulge
are also  released \citep{2010AcA....60..165S,2011AcA....61....1S} and some
work is also done with those data \citep{2011arXiv1107.3152P}. It is worth to
note that these work mainly deals with the structure of LMC other than the RR
Lyr variables.

In this paper, we focus the study on the RR Lyr variables themselves based on the
released OGLE-III database. With the PDM and Fourier fitting methods, we make a
precise systematic frequency analysis of carefully selected 655 RRLS  in LMC,
and a detailed classification of the RRLS in the LMC based on which to discuss the
incidence rate of the Blazhko modulation in various pulsation modes. The data
and the sample are illustrated in section 2,  the method is introduced in section
3,  the detailed classification of RRLS in LMC in section 4 and  discussion in
section 5.

\section{The sample}

\cite{2009AcA....59....1S} presented a catalog of 24,906 RR Lyrae stars
discovered in LMC based on the OGLE-III observations and classified them into
17,693 fundamental-mode, 4958 first-overtone, 986 double-mode and 1269 suspected
second-overtone RRLS. Thanks to their generosity, all the data are released.
This catalog has three columns recording the observational Julian Date,
magnitude in the I or V band and error of the magnitude. Since the number of
measurements in the V band is much fewer than in the I band, our analysis mainly
makes use of the I-band data. For over 20,000 objects, precise analysis of the
frequency for all of them seems an improbable task. Fortunately, the statistical
properties can be reflected by a much smaller sample. Thus, we concentrate our
study on a sample of RRLS that were measured as many times as possible with high
precision.

\begin{figure}
   \includegraphics[width=\textwidth]{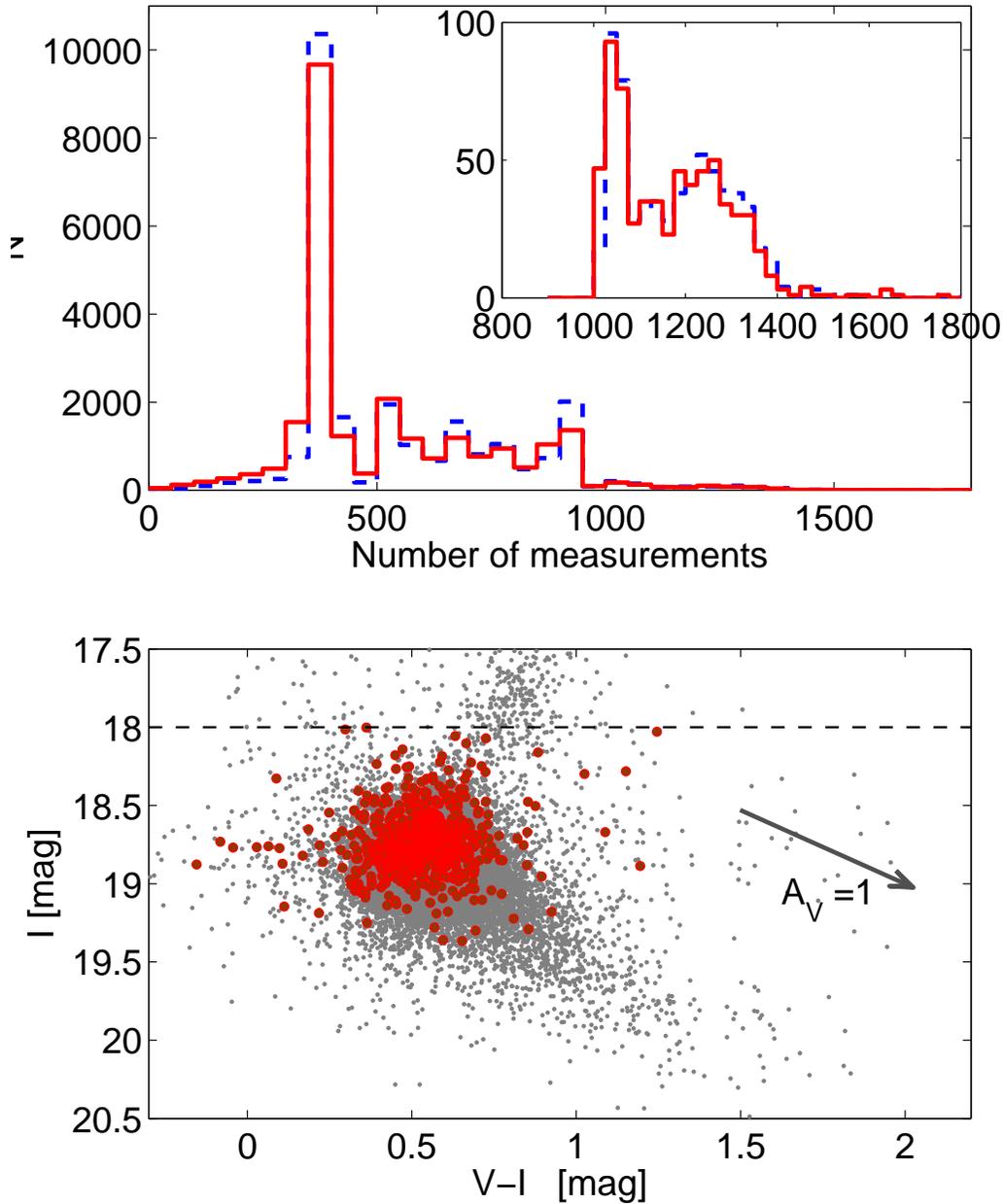}
\caption{Histogram of the number of measurements in the I band  and the color-magnitude diagram  of the OGLE III RRLS. Upper
panel: the blue dash line is for the data originally from the catalog and the
red solid line for those with photometric uncertainty smaller than 0.1 mag which
is the threshold when selecting our sample; the inset is the histogram of our
sample in which the photometric uncertainty is less than 0.1 mag and the number
of measurements is more than 1000. Lower panel:  the gray dots and red dots
correspond to all the RRLS from OGLE III and those in our sample
respectively; the dashed horizontal line at $\bar{I} = 18$ mag is our criterion
for brightness. The arrow stands for the $\rm{A_V}=1$ vector. }
\label{fig:1}
\end{figure}

As the amplitude of some RRLS is rather small as about 0.1 mag, the measurements
with assigned photometric error bigger than 0.1 mag are dropped. In the released
database, the number of measurements is usually not as numerous as claimed in
the OGLE-III catalog web. Most of them have fewer than 400 measurements, as can
be seen in Fig.~\ref{fig:1} that displays the distribution of the number of
measurements for all the RRLS. In our sample, only the objects with more than
1000 measurements are kept. To exclude the foreground stars,  the criterion
$\bar{I} \ge 18$ mag is added. In the I/V-I diagram (Fig.~\ref{fig:1}), it
can be seen that this brightness cutoff constrains the sources in the major RRLS
area and excludes some sparse sources seemingly to be giant or foreground stars.
 With the limitation of brightness and number of measurements, our sample
consists of 655 RRLS.

Through the color-magnitude diagram of the 655 RRLS in comparison with the
complete group of all the 24906 RRLS sources identified by
\cite{2009AcA....59....1S}  in Fig.~\ref{fig:1} (top panel), we can see that
the sample agrees with the majority of RRLS. The miss of the faint
sources (I $>$ 19.5) is mainly due to our request of high photometry quality. In
the same time, some red RRLS with V-I bigger than about 0.8 are neither included.
The faint and red RRLS may be caused by extinction as they coincide pretty well
with the A$_{\rm V}$=1 trend (the extinction law is taken from \cite{1990ARA&A..28...37M}) in Fig.~\ref{fig:1}. Thus the missed faint and red stars should have
intrinsically similar brightness and color with the majority, and their absence
in our sample shall not influence the statistical variation properties of RRLS.
This sample of 655 RRLS has three advantages for the frequency analysis.
Firstly, more than six-hundred stars are already big enough to obtain the
statistical parameters objectively and to understand the common properties of
RRLS in LMC. Secondly, the sample is not too big, so that we can not only make
the frequency analysis accurately, but also check carefully for individual
object, to make sure every result is reliable. Thirdly, the sample we compose
contains the highest-quality data for RRLS in the OGLE-III project, and the
derived variability properties should be highly reliable. As mentioned earlier,
all stars in our sample have more than 1000 measurements. The time span is about
4000 days (i.e. close to 11 years), and the interval of two adjacent
measurements is mostly shorter than 20 days. Theoretically, based on the effect
of random and uncorrected noise, using the least square fit of a sinusoidal
signal, we can roughly estimate the error of the frequency  $10^{-7}$, and the
error of amplitude $10^{-3}$ \citep{1999DSSN...13...28M}. The objects in our
sample are listed in Table~\ref{tab1}, with the OGLE name, position, number of
measurements both in original OGLE catalog and in our calculation, the magnitude in the I and V bands that are the average over all the measurements
, the subtype of variation which will be discussed
later, and MACHO ID if available. It should be noticed that the I magnitudes in the tables afterwards  are the average value of the Fourier fitting.

\begin{table}

\caption{List of the RRLS in our sample. The full table is available in electronic form at the CDS.}
\centering 
\begin{tabular}{c c c c c c c c c c}  
\hline\hline                        
ID	&	OGLE name	&	RA
&	Dec	&	$N_{ori.}$	&
$N_{cor.}$	&	$\bar{I}$	&	$\bar{V}$
&	subtype	&	MACHO ID	\\
\hline  
1	&	OGLE-LMC-RRLYR-00762	&	04:45:14.61	&
-68:21:28.2	&	1023	&	 1016	&	18.831	&	19.560
&	RR0	&	…	\\
2	&	OGLE-LMC-RRLYR-05348	&	05:07:31.21	&
-69:11:05.3	&	1179	&	 1176	&	18.549	&	18.948
&	RR1	&	1.4412.1128	\\
3	&	OGLE-LMC-RRLYR-05901	&	05:09:04.43	&
-69:36:07.0	&	1030	&	 1009	&	18.825	&	19.341
&	RR01	&	…	\\
4	&	OGLE-LMC-RRLYR-05976	&	05:09:15.63	&
-69:18:48.1	&	1035	&	 1033	&	18.701	&	19.347
&	RR0	&	1.4652.1639	\\
5	&	OGLE-LMC-RRLYR-05990	&	05:09:17.09	&
-69:21:13.8	&	1045	&	 1031	&	18.876	&	19.406
&	RR0	&	5.4652.8925	\\
6	&	OGLE-LMC-RRLYR-06004	&	05:09:19.04	&
-69:28:47.5	&	1035	&	 1028	&	18.723	&	19.240
&	RR0	&	…	\\
7	&	OGLE-LMC-RRLYR-06033	&	05:09:23.61	&
-69:31:17.8	&	1010	&	 1010	&	18.056	&	18.691
&	RR0	&	…	\\
8	&	OGLE-LMC-RRLYR-06114	&	05:09:34.62	&
-68:43:20.1	&	1029	&	 1027	&	18.788	&	19.448
&	RR0	&	2.4782.3916	\\
9	&	OGLE-LMC-RRLYR-06125	&	05:09:36.66	&
-68:55:30.5	&	1032	&	 1030	&	18.457	&	18.846
&	RR0	&	79.4779.1340	\\
10	&	OGLE-LMC-RRLYR-06147	&	05:09:39.41	&
-68:55:07.4	&	1017	&	 1009	&	18.793	&	19.380
&	RR0	&	79.4779.2347	\\
\hline  
\end{tabular}
\label{tab1}
\end{table}

\section{Frequency analysis}

Most studies of the RRLS frequencies (e.g. \citealt{2010ApJ...713L.198K}) make use
of the Period04 software that analyze the Fourier spectrum of the observed light
curve. The advantage of Period04 is its high precision and intuitive power
spectrum with both the frequency and amplitude shown, appropriate for analyzing
multi-period light-curves. But, the Fourier analysis fits a sinusoidal signal,
while for RRLS in the fundamental mode, their light-curves are very asymmetric.
Besides, RRLS often have several harmonics, so that more than one period would
be found for a single period RRLS. An alternative method to determine the
frequency of light variation is the Phase Dispersion Method (PDM,
\citealt{1978ApJ...224..953S}) independent on the shape of the light curve or
irregular distribution of measurements in the time domain. Although PDM also
brings about a strong signal of the harmonics, this can be looked over in the
folded phase curve, which needs a careful eye-check. The mediate volume of our
sample makes it possible to check the phase curve one by one. Moreover, the
result from PDM provides a mutual check between the two main-stream methods in
the study of variable stars.

The PDM method looks for the right period of light variation in a range of trial
periods by fixing the period of the minimum phase dispersion for the folded
phase curve. The phase dispersion $\Theta_{\rm PDM}$ is defined as the ratio
between the summed phase dispersion in all the phase bins to the phase
dispersion of all the measurements. A perfect periodical light curve would
produce $\Theta_{\rm PDM}=0$. The period of light variation of the RRLS sample
is searched in two steps. In the first step, the frequency range is set to the
whole range for RRLS variation, i.e. from 1 c/d to 5 c/d, with a step of
$10^{-5}$ and 50 bins in the phase space [0,1.0]. The PDM analysis
yields the first guessed frequency ($f_{\rm PDM}$) at the minimum $\Theta_{\rm
PDM}$. With this frequency, the folded phase light curve is plotted and checked
by eyes to exclude the harmonics, often the double or triple, which results in a
crude estimation of the main frequency, $f_{\rm est}=f_{\rm PDM}*n$ (n=1,2,3...
according to the order of the harmonics in the phased light-curves). In the
second step, the frequency resolution is increased to $10^{-7}$ and the range of
frequency is shrunk to [$f_{\rm est}$-0.05, $f_{\rm est}$+0.05]. Then a PDM
analysis is performed once more to yield the minimum $\Theta_{\rm PDM}$ that
tells the frequency with higher accuracy. Thanks to the long time span and
numerous measurements of RRLS in the sample, such high precision is achievable
in determining the frequency. This frequency is the main pulsation frequency,
$f_{0}$ or corresponding period $P_{0}$ in following text, to distinguish among
the fundamental, first overtone and second overtone modes.

Once the main frequency is determined, the light curve is fitted by a linear
superposition of its harmonic oscillations:
\begin{equation}
M(t)=M_{0}+\sum_{k=1}^{n} (a_k \sin{k \omega t}+b_k \cos{k \omega t}),
\end{equation}
where $n$ is the highest degree of the harmonics, $M(t)$ the measured magnitude in the I or
V band, $M_{0}$ the mean magnitude, and $\omega =2 \pi /P_{0}$ the circular
frequency. In fact, it's the phased light-curve instead of the light-curve
itself that's fitted for the sake of higher significance:
\begin{equation}
M(t)=M_{0}+\sum_{k=1}^{n} (a_k \sin{2 \pi k \Phi _t}+b_k \cos{2 \pi k \Phi _t}),
\end{equation}
where $\Phi_{t}=(t-t_0)/P-|(t-t_0)/P|$, $t$ is the time of observation and $t_0$
is the epoch of maximum brightness. Eq. (2) can be re-written as:
\begin{equation}
M(t)=M_{0}+\sum_{k=1}^{n} A_k \sin{(2 \pi k \Phi _t+\phi _k)},
\end{equation}
where $A_{k}=\sqrt{a_k^{2}+b_k^{2}}$ and $\phi_{k}=\arctan{(b_k/a_k)}$. The parameters  $A_k$ and $\phi_k$  can be transformed to the Fourier parameters
$R_{ij}=A_j/A_i$ and $\phi_{ij}=j\phi_i-i\phi_j$ ($i$, $j$ refers to different
$k$, \citep{1981ApJ...248..291S}), both of which are widely used in expressing
the features of the light-curves, and even to derive the physical
parameters of the variables such as the metallicity (e.g.
\citealt{1996A&A...312..111J}).

\begin{figure}
   \includegraphics[width=\textwidth]{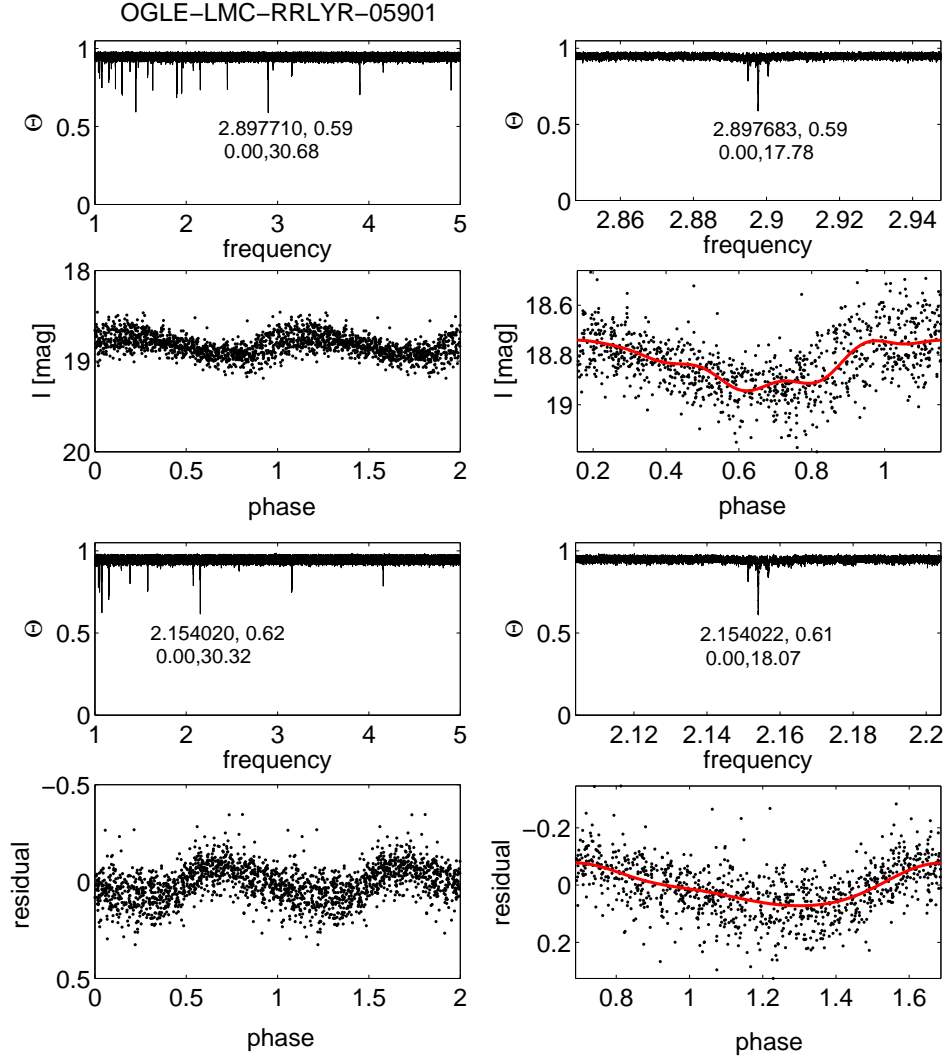}
\caption{An example of the frequency identification (for Star 3): the first loop
to search for the primary frequency from original data (upper 4 figures) and the
second loop to search for secondary period from the residual data after first
prewhitening (lower four figures). For either loop of the four figures,
the left two figures are $\Theta _{\rm PDM}$-frequency diagram with
frequency in [0.2, 1, 0.00001], phased light-curve based on the frequency
$f_{\rm 1st}$, while the right two figures are $\Theta _{\rm
PDM}$-frequency diagram with frequency in [$f_{\rm est}$-0.05, $f_{\rm
est}$+0.05, 0.0000001], and phased light-curves based on the frequency $f_{\rm
final}$ where the red line is the Fourier fitting to the phased light-curve. The
numbers shown in the $\Theta _{\rm PDM}$-frequency figure are the frequency
derived and corresponding $\Theta_{\rm PDM}$, $P$ and $S/N$ described in Section
3. }
\label{fig:2}
\end{figure}

In principle, the degree of the harmonics can be arbitrarily high, however, the
highest degree in practice is set to 5 or 6, being able to reflect the essential
shape of the light curve. Among all the 655 stars, only 64 stars need the sixth
harmonics and all the others with $n\leq$5. Moreover, it avoids
over-fitting, even for very complex light curves such as shown in
Fig.~\ref{fig:2}. The main pulsation frequency we derived by this method is
almost the same as that of \cite{2009AcA....59....1S}, with the difference
$\leq$ 0.0001. The upper four panels in Fig.~\ref{fig:2} show the process of
determining the primary frequency, from the estimation of the frequency, through
the accurate measurement of the frequency and the folded phase curve to the
final fitting of the phase curve.

The secondary period is searched in the residual after subtracting the variation
in the main frequency with its harmonics. The method is the same as for the
principle period. The difference lies in the amplitude of the secondary oscillation
being much smaller than the principle one. This procedure is carried repeatedly
until an assigned threshold, which is set to guarantee the significance of the
derived period. The phase dispersion parameter $\Theta$ is related to the
significance of the period, the smaller the better. Another related parameter is
the probability $P$ ($F(P/2,N-1,\sum(n_j)-M)=1/\Theta)$
\citep{1978ApJ...224..953S}. However, $\Theta$ and $P$ depends on the number and
quality of measurements so they do not necessarily have the same cut-off value
in different cases  and  become ambiguous in marginal cases. We developed an
independent parameter, $ S/N\equiv
\frac{\bar{\Theta}-\min(\Theta)}{\sigma_{\Theta}}$, which reflects the S/N of
the minimum $\Theta$ in the $\Theta$ distribution. Combining all the three
parameters, the specific thresholds are set to $\Theta_{\rm PDM}<0.63-0.85$,
$P<0.01-0.05$ or $S/N >10-15$ for a reliable period determination. Once the
period is determined, a fitting is performed as for the principle period, but
the highest degree of the harmonics is taken to be 3 instead of 6 in the first
run. The lower four panels in Fig.~\ref{fig:2} show the procedure for
determining a secondary period, including the first guess and final
determination of the frequency and the fitting of the phased light curve.

\section{Classification}

RRLS are considered as  the pure radial pulsator basically.
According to the pulsation modes, it was classically divided into four types:
RRab pulsating in the fundamental (FU) mode, RRc in the first overtone (FO)
mode, RRe in the second overtone mode (SO) and RRd in the double (FU and FO)
modes. \cite{2000ApJ...542..257A} introduced a new system of notation with a
digit to mark the primary pulsation mode to replace the letters, i.e.  RR0, RR1, RR2
and RR01 instead of RRab, RRc, RRe and RRd, more intuitive to mnemonics. When the
modulation of period and amplitude is considered,  \cite{2006A&A...454..257N}
adopted additional letters to classify RRLS, PC for period change, BL for the
Blazhko effect and MC for  closely spaced multiple frequency components.  To
make a complete view of the variation type, we combine both notations into a
more detailed designation to make the phenomenological classification of RRLS by
following \cite{2006A&A...454..257N}.

\begin{figure}
   \includegraphics[width=\textwidth]{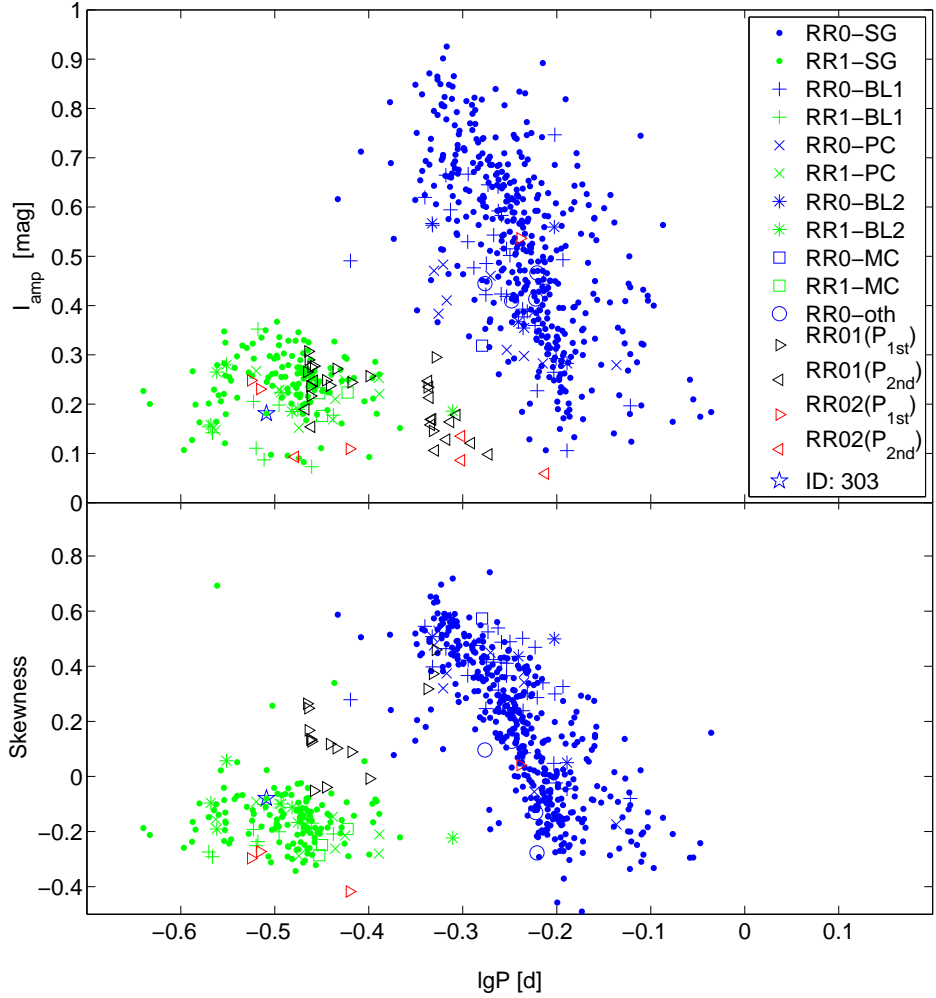}
\caption{The period-amplitude and period-skewness diagrams for all RRLS in our
sample. The definition of skewness is given in the text. The meanings
of the symbols are shown in the legend panel where P$_{\rm 1st}$ and P$_{\rm
2nd}$ means the primary and secondary period respectively in double-mode RRLS,
and the meaning of specific classifications is described in text in Section 4.}
\label{fig:3}
\end{figure}

\begin{figure}
   \includegraphics[width=\textwidth]{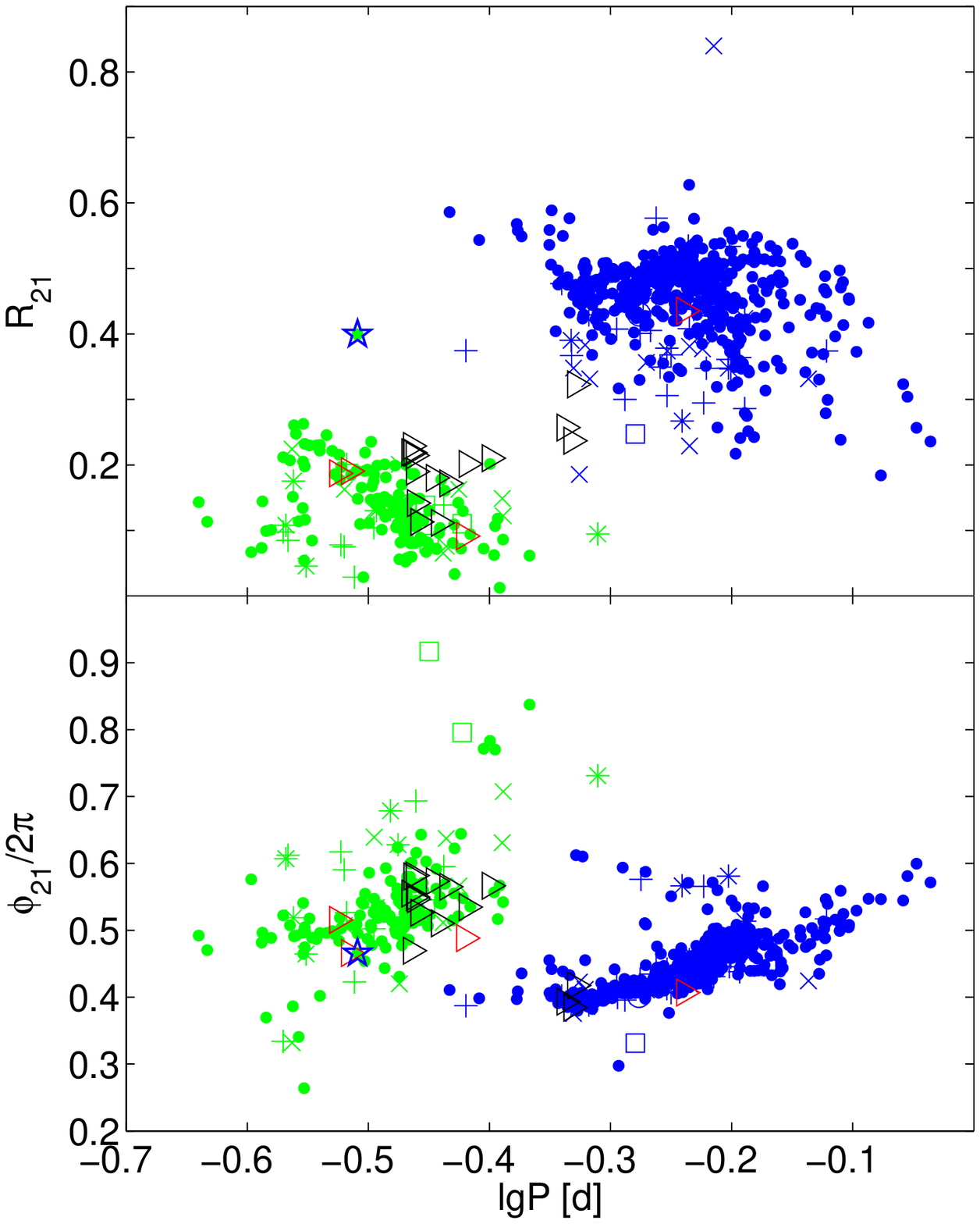}
\caption{The period-Fourier coefficients diagram. The symbols obey the same
legend as in Fig.~\ref{fig:3}. The Fourier coefficients are defined
in Eq.(3) with $\phi$ divided by $2\pi$.} 
\label{fig:4}
\end{figure}

The identification of the main pulsation mode is clear for separating the RR0
and RR1 classes, as proved in many previous studies. The period is longer and
the amplitude is mostly larger in RR0 RRLS than in RR1, and the shape of light
curve of RR0-type RRLS is more asymmetric than the RR1-type, as shown in the
period-amplitude and period-skewness diagrams in Fig.~\ref{fig:3},  where the
skewness  is calculated from the phased light curve. The definition of
skewness is $E(x- \mu) ^3 / \sigma ^3$, where x is the observed magnitude, $\mu$  the mean of x, $\sigma$  the standard deviation of x, and $E(t)$
represents the expected value of the quantity $t$.  Examples of phased
light-curves in our sample are shown in Fig.~\ref{fig:5}.  The gap between RR0
and RR1 is also clearly shown in the period-Fourier coefficients diagram in
Fig.~\ref{fig:4}.

The puzzle comes with the identification of the RR2-type RRLS, those pulsating
in the second-overtone mode. The RR2 stars are believed to have even shorter
period, slightly smaller amplitude and more symmetric light curve. In the
period-amplitude diagram, they should locate on the left of RR1 stars, e.g. the
magenta points in Fig.2 of \cite{2009AcA....59....1S}. In the period-amplitude
diagram  of our sample (Fig.~\ref{fig:3}, top), no clear gap is found in the
shorter-period group of stars. There is neither apparent peak in the period
distribution of all single-mode RRLS  shown in Fig.~\ref{fig:6}, which is very
similar to that of all the RRLS in LMC \citep{2009AcA....59....1S}. Concerning
the shape of the light curves, their skewness is calculated. From the appearance
in the bottom panel of Fig.~\ref{fig:3}, the separation between RR0 and RR1 is
again apparent with RR1 being systematically more symmetric, while no more
subgroup can be further distinguished in the relatively symmetric group. Thus, if a RR2
group is to be assigned, the borderline would be very arbitrary both in the
period-amplitude and period-skewness diagrams. Because there is no systematic
features, we are conserved in identifying a group of RR2 stars. In fact, it's
also possible that the stars with shorter period, smaller amplitude and more
sinusoidal light curve may be metal-rich RR1 stars \citep{1997ApJ...483..811B}.

\subsection{Single period RRLS}

The notation ''RR-SG'' refers to the single period RRLS, i.e. no more frequency
is found in the residual after removing the primary frequency and its harmonics.
In our sample, 556 stars were found to be RR-SG stars,  84.9\% of all the 655
sample RRLS. Out of these RR-SG stars, 424 (76\%) are RR0-SG in the FU mode and
132 (24\%) are RR1-SG RRLS in the FO mode. The RR0-SG RRLS are three times as
many as RR1-SG RRLS. In Table~\ref{tab2} we listed their period of light variation,
minimum phase dispersion $\Theta$, amplitude and mean magnitude in the I band,
and subtype. The histogram of the RRLS periods (Fig. ~\ref{fig:6}) displays two
classical prominent peaks at 0.58 d and 0.34 d for RR0 and RR1 stars
respectively. The average period of our sample for RR0 stars  $\bar{P}_{\rm
RR0-SG}$ is 0.587~d and for RR1 stars $\bar{P}_{\rm RR1-SG}$ is 0.328~d .

\begin{figure}
   \includegraphics[width=\textwidth]{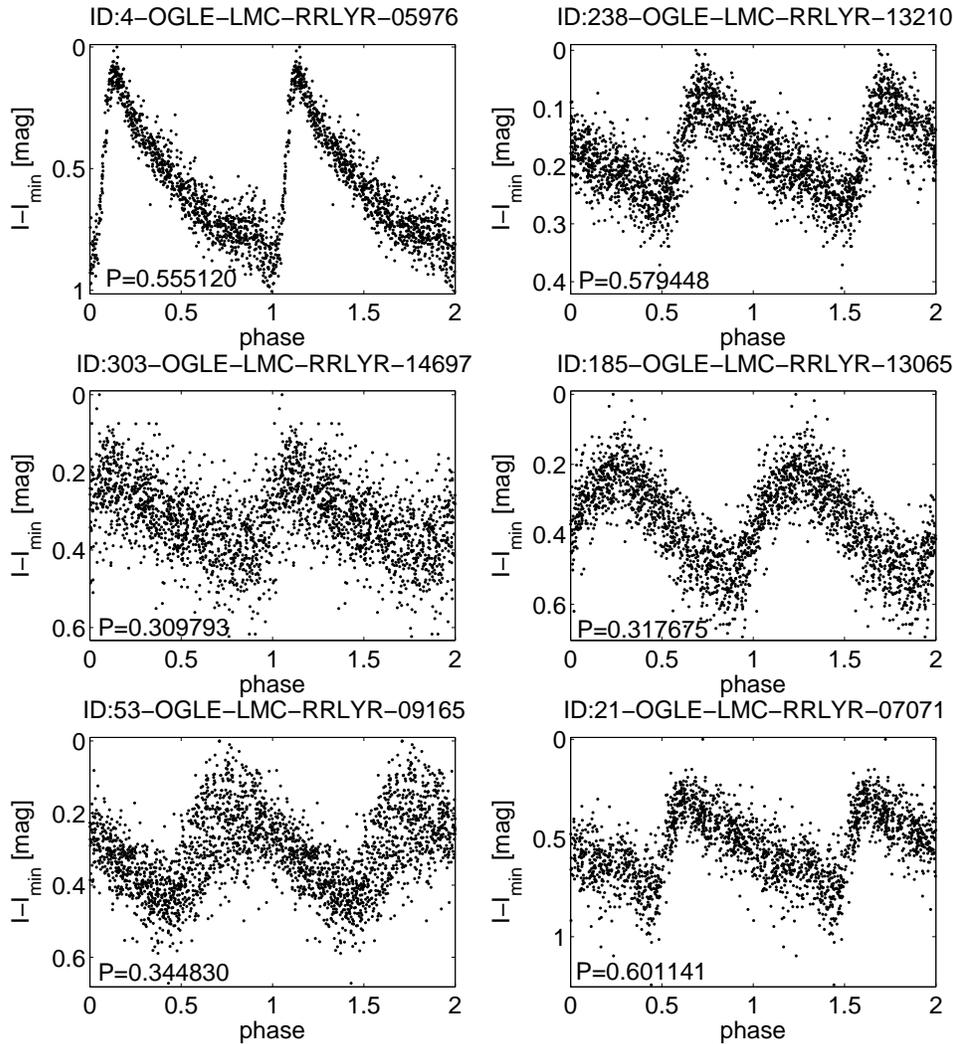}
\caption{Examples of the phased light-curves, including the controversial star
ID303 (middle-left).  The upper two stars are of type RR0 , the middle two of
RR1 and the lower two of RR01 (left) and RR02 (right) respectively.}
\label{fig:5}
\end{figure}

\begin{figure}
   \includegraphics[width=\textwidth]{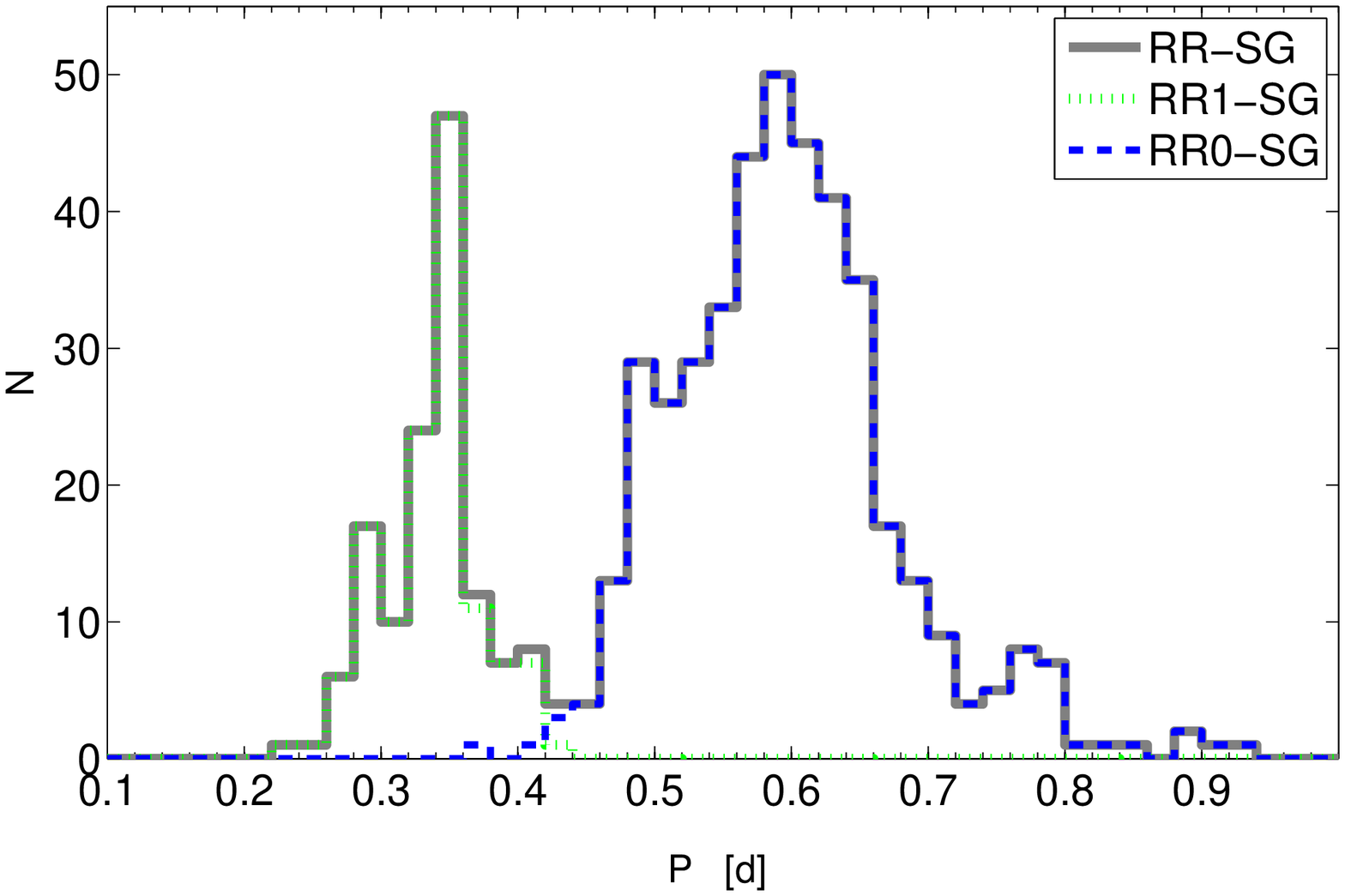}
\caption{Period distribution of all the single-mode RR-SG stars, with blue dash
line for RR0-SG, green dash line for RR1-SG, and gray solid line for the sum of
RR0-SG and RR1-SG RRLS.}
\label{fig:6}
\end{figure}

We compare our division of the subtype into RR0 and RR1 with that of
\cite{2009AcA....59....1S}, regardless of RR2 stars. One star (ID: 303, OGLE
ID: OGLE-LMC-RRLYR-14697) is found to be discrepant, which is classified as RR1
in our work and RR0 in their work. In Fig.~\ref{fig:3} that is the main criteria
for classification, this star is specially denoted by a pentagon. In either the
period-amplitude or the period-skewness diagram, this star locates in the
central area of RR1 stars. When compared with the RR0 and RR1 stars in the
phased light curve (Fig.~\ref{fig:5}), its shape is apparently asymmetric, but
not the same as RR0 stars (not so steep as RR0 stars). On the other hand, its
small amplitude and short period bring it to the RR1 group. This example also
tells that the skewness of the light curve of one class covers a wide range, and
makes the separation of RR2 stars hard.

\begin{figure}
   \includegraphics[width=\textwidth]{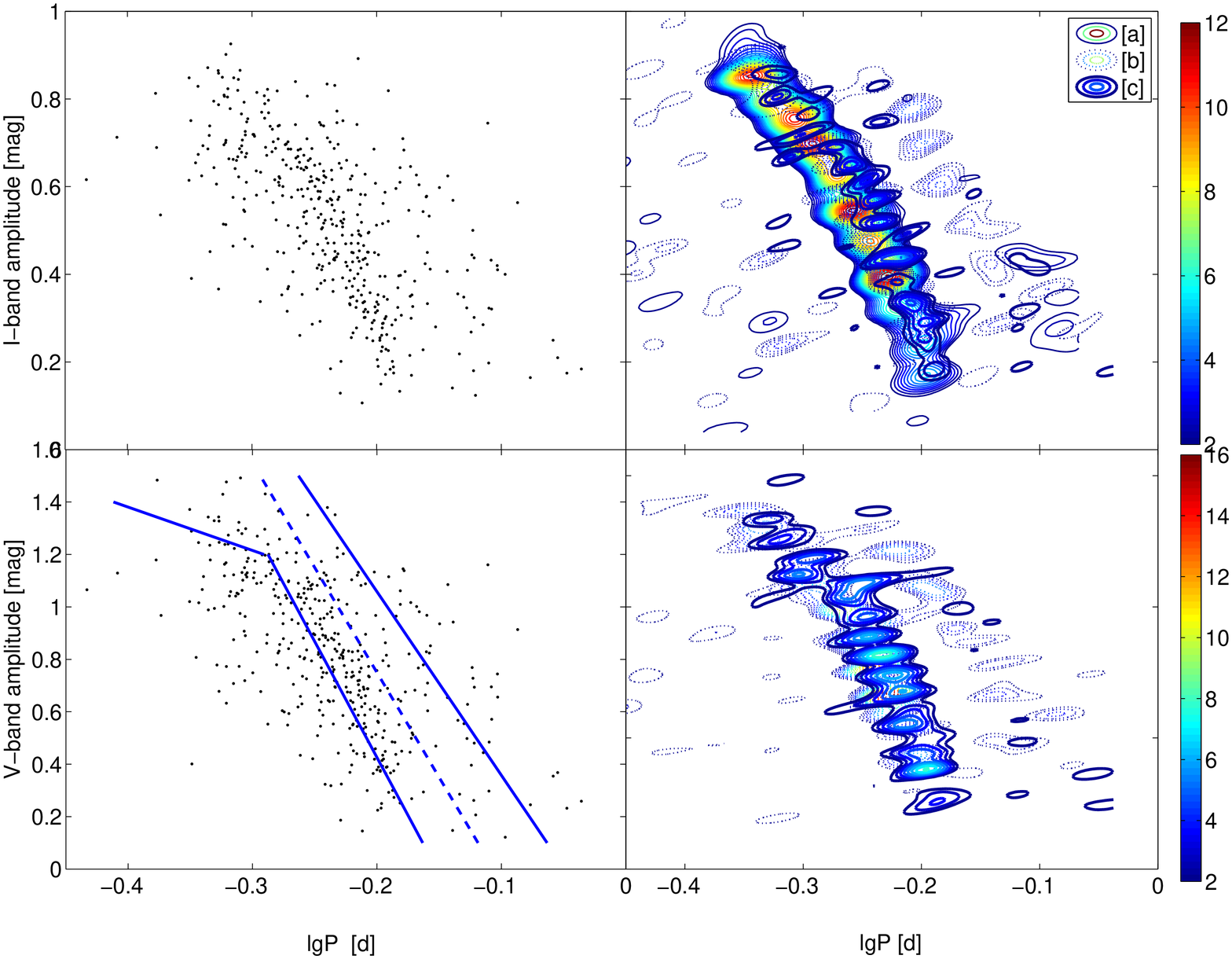}
\caption{Period-amplitude (I-band and V-band) diagram of RR0-SG stars. Contours
show the distribution of the bulge RR0-SG stars from [a]:
\citet{2006ApJ...651..197C} and [b]: \citet{2009AcA....59..137S} as well as [c]
of the LMC RR0-SG stars in this work. The two solid lines in the V-band diagram
are from \citet{2009AcA....59..137S}, showing the fitting of the Bailey diagram
for the Oo I and II RRLS respectively, while the dashed line is the border to
separate the Oo I type and Oo II type RRLS in the Galactic RR0 stars.}
\label{fig:7}
\end{figure}

The distribution of RR-SG in Fig.~\ref{fig:3} is composed of two typical parts,
one sequence of RR0-SG and the other shape of bell for RR1-SG. The distribution
of RR0-SG seems to be composed of two parts, one densely clumped on the left
forms a line shape, the other loosely distributed on the right. This shape of
distribution reminds one of the dichotomy into Oosterhoff I and II classes of
Galactic RRLS due to different evolving phases. Such dichotomy was found in the
Galactic fundamental-mode RRLS \citep{2009AcA....59..137S}. Fig.~\ref{fig:7}
compares the Bailey diagram of RR0-SG stars in LMC with that in the Galactic
bulge \citep{2006ApJ...651..197C} and the Galactic field
\citep{2009AcA....59..137S}, where the contours are for the density  of RR0-SG
stars, the solid and dashed lines are from Equation 2, 3, 4 and 5 of
\cite{2009AcA....59..137S} for their fitting to the  Oo I (left), Oo II (right)
groups and their borderlines respectively in the Galactic field RRLS. It can be
seen that the RR0 stars in LMC can be divided into the Oo I and Oo II groups as well as 
in the Galactic field and bulge. Although the two groups are not fitted,  the
solid and dash lines from \cite{2009AcA....59..137S} generally agree with the
distribution. Moreover, the Oo I group RR0 stars are clearly dominating over the
Oo II stars. This is consistent with the fact that the average period
$\bar{P}_{\rm RR0-SG}$=0.587~d is more approximate to 0.549~d of the average Oo I clusters
than to 0.647~d of the average Oo II clusters \citep{2000AAS...196.4103C,
1959AnLei..21..253V}. According to the contour diagram, the Oo I group RR0 in
LMC stars have a slightly longer period ($\bigtriangleup \lg P \sim 0.02$) than
those in the bulge. If the bulge and field appearance in the contour diagram is
taken into account, the Oo I distribution forms a series from the bulge to the
field and then to the LMC from left to right, i.e. the period from short to
long. This sequence coincides with that of the average metallicity in these
three environments. Since metallicity influences the opacity and the mechanism
for the light variation of RRLS is the $\kappa$ mechanism, the shift of the Oo I
group may be caused by the metallicity difference. Indeed, the metal abundance
also influences the distribution of OoI and Oo II RR0 stars in the Galactic
field \citep{2009AcA....59..137S}.

\subsection{Multiple period RRLS}

There are 99 (15.1\%) RRLS with variation detected in the residual of the light
curve after removing the principle frequency and its harmonics. They are further
classified into several subclasses according to the number of additional
frequencies and their locations relative to the main frequency, including the
RR01, RR-BL, RR-MC, RR-PC and miscellaneous subtypes.

\subsubsection{RR01 stars}

RR01 refers to the RR Lyrae stars pulsating in double radial modes, one is the
FU mode and the other is the FO mode, classically RRd stars. In our sample, 15
($2.3\%$ in the sample) stars are classified as RR01 stars. Seven of them have
two frequencies detected. Other eight stars have three frequencies detected, but
the third frequency is either the sum or the difference of the first and second
frequency and thus dependent. All these 15 stars are listed in Table~\ref{tab3}, with the
period and amplitude in the FU mode P$_{0}$ and A$_{0}$, the minimum phase
dispersion $\Theta_{0}$ in deriving the FU mode, the mean magnitude in the I
band, the period and amplitude ratios between  FO and FU
modes, and the minimum phase dispersion $\Theta_{1}$ in deriving the FO
frequency.

In the period-magnitude diagram (Fig.~\ref{fig:3}, top), both the FU and FO
periods of these double mode RRLS are shown by black open triangles. The periods
are either at the long side of the single FO mode or the short side of the FU
RRLS. Their amplitudes are small, all smaller than 0.3~mag, which is comparable
to that of the RR1 stars. The dominance of FO mode in RR01 can explain this
small amplitude. It can be seen that the amplitude of the FU mode of RR01 stars
is much smaller than those single mode FU stars at corresponding short period
while comparable to those single mode FO stars. In fact, RR01 stars distinguish
themselves from the other single-mode FU stars by their small amplitude in the
period-amplitude diagram. Meanwhile, their light curves appear differently by a
positive skewness from the FO RRLS most of which have a negative skewness
although both are more symmetric than the FU RRLS. The lower panel of
Fig.~\ref{fig:3} shows such difference.

\begin{figure}
   \includegraphics[width=\textwidth]{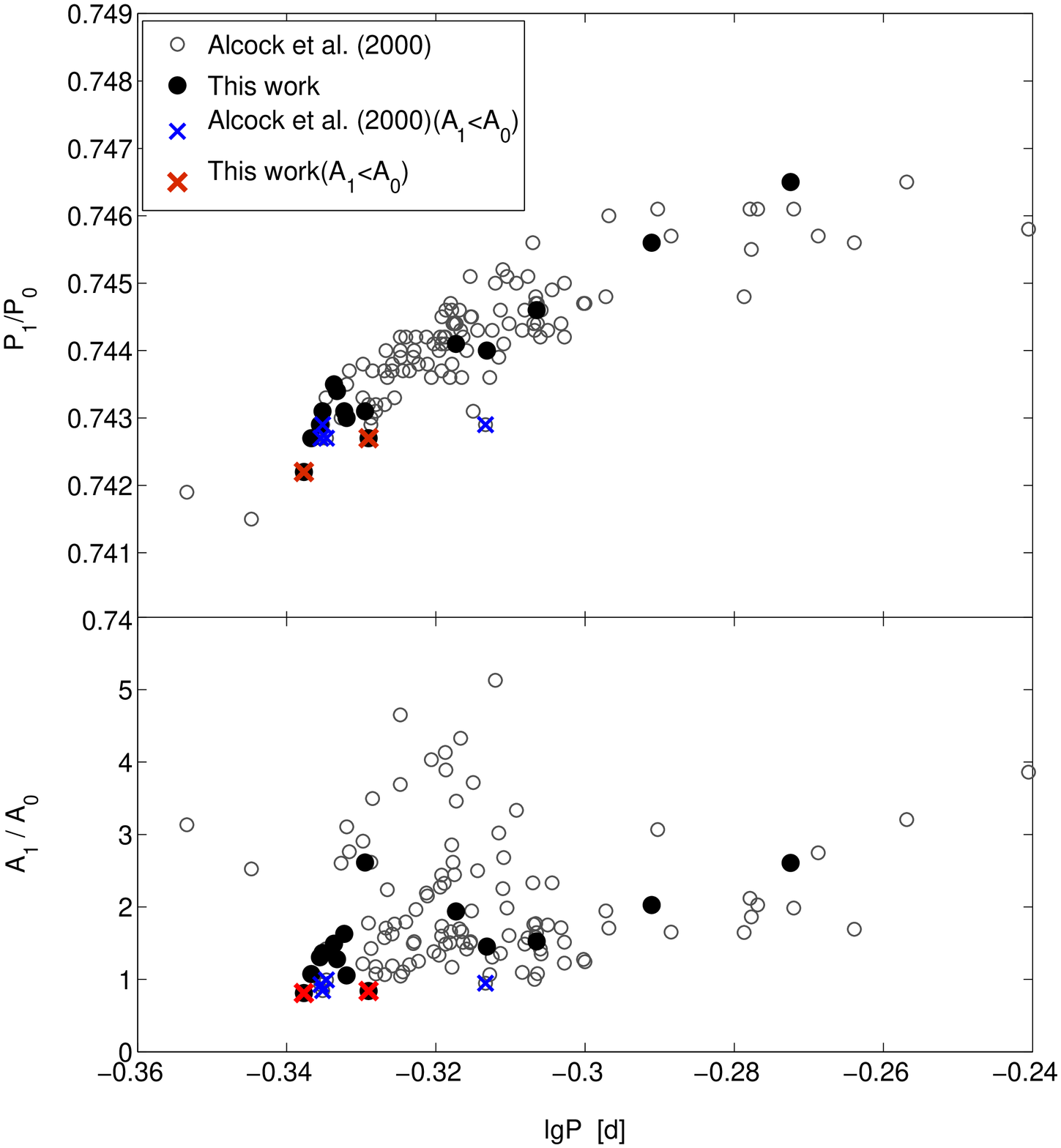}
\caption{Petersen's diagram for the RR01 stars (upper) and the amplitude ratio
(lower) vs. $\lg$P diagram in LMC. The solid symbols are from our results while the
hollow ones from the LMC data by the MACHO project \citep{2000ApJ...542..257A}.
}
\label{fig:8}
\end{figure}

The period ratio $P_1/P_0$ ranges from 0.7422 to 0.7465, with an average of
0.7436. The amplitude ratio $A_1/A_0$ is significantly larger than one in 13
stars, and two smaller than one (but bigger than 0.8), with an average of 1.533.
 It can be concluded that the RR01 stars mainly pulsate in the FO mode. In the
Petersen diagram for RRLS (Fig.~\ref{fig:8}), the period and amplitude ratios
are plotted versus the FU period, and compared with the results of
\cite{2000ApJ...542..257A} from the MACHO data. The increase of both ratios
with the period is clear and consistent with \cite{2000ApJ...542..257A}. The
distribution of the period ratio overlaps completely with that of
\cite{2000ApJ...542..257A}. The ratio of the amplitude does not rise so high as
the \cite{2000ApJ...542..257A} result, although the rising tendency with period
is the same.  Moreover, the RR01 stars whose dominating mode is fundamental
(i.e. $A_1/A_0$ $<1$) have smaller  $P_1/P_0$ than the average, specifically,
their $P_1/P_0$ is all smaller than 0.743, even when including those from
\cite{2000ApJ...542..257A}.

\subsubsection{RR-BL stars}

RR-BL stars refer to the Blazhko stars. As mentioned in Introduction, the early identification of RR-BL stars was the symmetric appearance of the frequency spectrum. With the development of the study of the Balzhko effect, some asymmetric patterns are considered to be its variation. Thus, in the frequency pattern, they may appear as: (1) two
frequencies which have one closely spaced frequency around the
principle frequency (RR-BL1), (2) three frequencies which have two side frequencies closely and symmetrically distributed
around the main frequency (RR-BL2), and (3) more than three frequencies which have multiple components at closely spaced frequencies (RR-MC).
All of them are the consequence of the modulation of the amplitude and/or phase, which can be explained by the modulation
of a single (sinusoidal or non-sinusoidal) oscillation \citep{2009CoAst.160...17S, 2011arXiv1106.4914B}.

\paragraph{RR-BL1 stars}

RR-BL1 stars have one frequency close to the main frequency, bigger or smaller. \cite{2000ApJ...542..257A} marked them as $\nu$1, and here
we use the definition of  \cite{2006A&A...454..257N} to mark them as BL1.
There are 41 (6.3\% of all the 655 stars in the sample) such RR-BL1 stars,
forming a much larger group than other multi-period RRLS. Taking into account
the main pulsation mode, they are classified into 32 RR0-BL1 stars in the FU
mode and 9 RR1-BL1 stars in the FO mode. The main pulsation period of RR0-BL1
stars ranges from 0.38 d to 0.76 d and of RR1-BL1 from 0.26 d to 0.37 d.  The
main pulsation amplitude of RR0-BL1 stars ranges from 0.106 mag to 0.747 mag and
of RR1-BL1 from 0.073 mag to 0.352 mag.  In Fig.~\ref{fig:3} and
Fig.~\ref{fig:4}, these RR-BL1 stars are mixed homogeneously with other
single-mode RRLS, which means they have normal main period and amplitude of
light variation as those single-mode stars.

The sum of all the differences between every side frequency and main frequency
$\delta f=\sum_ i \Delta f_i$,  where $\Delta f_i =f_i-f_0$ and $i=1,2...$ for
frequency at the first, second overtone and so on , is usually used to
characterize the asymmetry of the frequency distribution. For RR-BL1 stars,
there is only one side frequency, resulting $\delta f = f_1-f_0$. About 75\% (24
out of 32) of these RR0-BL1 stars and 67 \% (6 out of 9) of these RR1-BL1 stars
have $\delta f$ positive. This is very different from the 37\% proportion for
RR1-BL1 stars derived from the MACHO data by \cite{2000ApJ...542..257A}. But it
agrees well with the percentage 80\% for RR0-BL1 stars from the study of
the Blazhko variables by \cite{2002ASPC..259..396K}.

For RR0-BL1 stars, the Blazhko periods vary between about 23 days and over 1500
days with an average of about 180 days and rms of 348 days. For RR1-BL1 stars,
the Blazhko periods vary from 6.4 days to about 3000 days with an average of 745
days and rms of 1404. The shortest modulation period (6.4 d) is comparable to
that found for RR1-BL in LMC by \cite{2006A&A...454..257N} and also consistent
with those in the Galactic field for RR0-BL stars \citep{2005AcA....55..303J},
i.e. around 6 days which is about 20 times of the main pulsation period. On the
other end, the longest modulation period $\sim$ 3000 d is comparable to the time
span of the data available, it is at least partly limited by the observational
time coverage. With the continuation of the OGLE project, longer modulation
period can be expected. The distribution of the Blazhko periods of RR-BL1 stars
is shown in Fig.~\ref{fig:10}. The RR0-BL1 stars exhibit a normal distribution
with the peak around 60-70 days.  The RR1-BL1 stars shows a quite scattering
distribution in a much wider range, although the group of only 9 RR1-BL1 objects
makes this statistical significance less convincible. It seems that there is a
preferred range of Blazhko period for RR0-BL1 from 32 d to 200 d ($\lg$P from
1.5 to 2.3), but no such preferred range for RR1-BL1 , which agrees well with
the result of \cite{2006A&A...454..257N}.

We listed the main frequency and other variation parameters of RR-BL1 stars in
Table~\ref{tab4}. The amplitude ratio ($A_1/A_0$) of RR0-BL1 varies from 0.119 to 0.382
with an average of 0.237 and of RR1-BL1 varies from 0.324 to 1.081 with an
average of 0.574.  RR1-BL1 have a larger amplitude ratio than RR0-BL1, with one
star (ID: 126) even larger than one but its main amplitude is small, only 0.087
mag.

\begin{figure}
   \includegraphics[width=\textwidth]{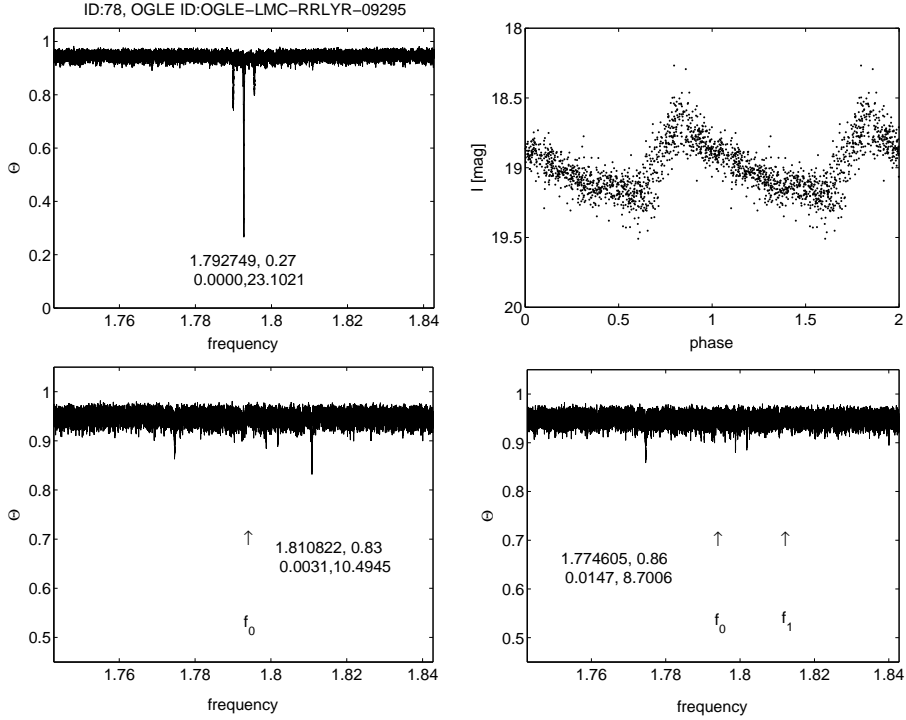} 
\caption{One example (Star 78) of the BL1 star which shows asymmetric
frequency as the result of the extreme amplitude asymmetry in the Blazhko effect
when the invisible component is completely submersed in the noise, where the
legends are the same as in Fig.~\ref{fig:2}. Notice that the
$\Theta$-Frequency diagram is different from the frequency spectrum, the
"triplets" shown in the first figure are only the main frequency and two aliases.}
\label{fig:9}
\end{figure}

\begin{figure}
   \includegraphics[width=\textwidth]{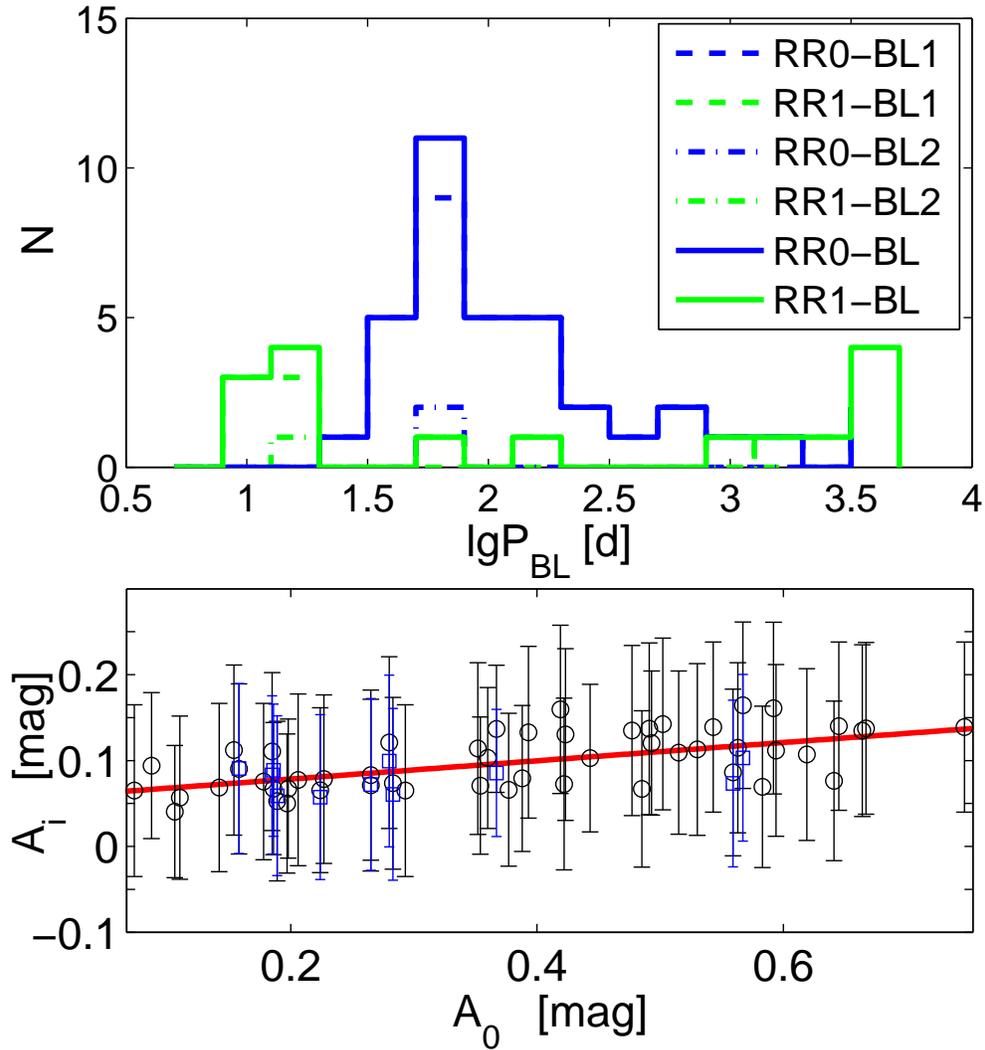}
\caption{Top: distribution of lg\rm{$P_{\rm BL}$} ($P_{\rm BL}$: the Blazhko
modulation period), where the meaning of symbols are explained in legends.
Bottom: the modulation amplitude vs. the main pulsation amplitude, where black
 filled circles denote the amplitudes of the second (in case of
RR-BL2) component of modulation, and the black solid line is the
linear fitting between the two parameters. }
\label{fig:10}
\end{figure}

The asymmetric frequency can be regarded as the extreme case of the
amplitude asymmetry in the Blazhko effect when the invisible symmetric component
is completely submersed in the noise.  In Fig.~\ref{fig:9} is shown an example
of such situation (star ID: 78, OGLE ID: OGLE-LMC-RRLYR-09295). The upper two
figures are the frequency-$\Theta_{\rm PDM}$ diagram and the phased light-curves
of the first-loop period searching for the main frequency 1.7927. The lower left
 figure shows the successful search for the secondary frequency at 1.8108, and
the lower right figure shows that no more reliable frequency can be derived
since all the three parameters at $f$=1.7746 ($\Theta_{\rm PDM}$=0.86,
sig.=0.015 and S/N=8.7) are below our threshold. On the other hand, it may be
expected that this frequency  would be detected given a higher sensitivity of
observation or a lower threshold. This example further supports that the missing
of another frequency component in RR1 stars be caused by the asymmetry of the
modulated amplitude.

\paragraph{RR-BL2 Stars}

For RR-BL2 stars, the secondary frequencies indicate the modulation of
the amplitude and phase. There are 11 (1.7\%)
RR-BL2 stars and they can be divided into two subtypes, 4 RR0-BL2 and 7 RR1-BL2
based on the main pulsation mode. This percentage (1.7\%) is much smaller than
the RR-BL1 stars (6.3\%). It is also low in comparison with the results of
previous studies, which will be discussed later.  We think such low percentage
is mainly due to our very strict criteria to identify a  frequency so that some
third frequencies have been dropped like the case of Star 78  shown in
Fig.~\ref{fig:9}. This 1.7\% percentage should be taken as the lower limit of
the percentage of RR-BL2 stars.

The differences between the side and main frequencies are shown in Table~\ref{tab5}, i.e.
$\bigtriangleup f_{+}$ and $\bigtriangleup f_{-}$. They are both smaller than
0.1. In addition, the difference between $\bigtriangleup f_{+}$ and
$\bigtriangleup f_{-}$  is  all smaller than 0.0003. It's these two features
that bring them into the RR-BL2 class.  On the ratio of the two amplitudes $A_+$
and $A_-$ ($A_+/A_-$), it changes from 0.76 to 1.60. This range of ratio means
the two components have pulsation amplitudes at the same order, or we are only
sensitive to such situation. This is understandable since a large ratio of the
two amplitudes would surely make the weak component invisible and move the star
into the RR-BL1 group. Such bias can only be alleviated by a very-high-sensitivity observation. This fact can also account for the low percentage of
the BL2 stars.

As shown in Fig.~\ref{fig:3} by the symbol asterisks, the RR-BL2 stars
have ordinary period and amplitude in the principle pulsation mode. The
amplitude ranges from 0.283~mag to 0.567~mag and the period from 0.465~d to
0.647~d, with the average $\overline{A_{0}}$= 0.44~mag and $\overline{P_{0}}$=0.58~d
for RR0-BL2. For RR1-BL2 stars, the amplitude ranges from 0.158~mag to 0.265~mag
and the period from 0.270~d to 0.489~d, with the average $\overline{A_{0}}$=
0.21~mag and $\overline{P_{0}}$=0.33~d.

According to $\bigtriangleup f_{+}$ and $\bigtriangleup f_{-}$, the modulation period
varies from about 43 to over 2700 days with an average of 1349 days for
RR0-BL2; and from 12 to 2902 days with an average of 1288.5 days for
RR1-BL2. The distribution of the modulation frequencies are shown in
Fig.~\ref{fig:10}. Because the volume of the RR-BL2 stars is small, the
distribution does not exhibit any outstanding feature. However, the
situation becomes clearer when the RR-BL1 stars are included, which is
reasonable since BL1 stars can be regarded as the extreme case of RR2 and both
are Blazhko variables. Consequently, our sample of 655 RRLS contains 52
Blazhko stars. In Fig.~\ref{fig:10},  the period distribution of all
the RR0-BL and RR1-BL stars is shown. Because of the dominance of RR-BL1 stars, the
distribution of RR-BL stars is similar to that of RR-BL1 stars, i.e. with a
preferred range of period from a few tens to a couple of hundred days. In
regards to the modulation amplitude, a correlation is found with the main
pulsation amplitude. As shown in Fig.~\ref{fig:10} (bottom), a linear fitting
results in that $A_{i}=0.106*A_{0}+0.057$ and the correlation coefficient is
0.605 which means significant correlation. The error here we adopted is the maximum of the photometric error assigned in the catalog which is apparently bigger than the error in the fitting. The order of the error is mostly around 0.1~mag. This correlation was not found before
and neither predicted in any models for the Blazhko effect. But it indicates
that the Blazhko modulation is related to the main pulsation mode and it should
be taken into account in models. On the contrary,
\cite{2005IBVS.5666....1J} found that the possible largest value of the
modulation amplitude, defined as the sum of the Fourier amplitudes of the first
four modulation frequency components, increases towards shorter period
variables.

\paragraph{RR-MC stars}

We have got 4 RR-MC in our sample.
Actually all of them have three side frequency
components. One RR-MC star is in the FU mode  and three are in the FO mode.
According to the structure of the side frequencies, they show three patterns,
similar to the RR-MC stars described in \cite{2006A&A...454..257N}: (a) two of
the three frequencies are symmetric to $f_{0}$; (b) none is symmetric to the
others, but three frequencies are at both sides to $f_{0}$; (c) three side
frequencies are all at one side to $f_{0}$. In Table~\ref{tab6}, the four RR-MC stars are
shown with their patterns in the column "Notes". These stars with
multiplets may also be  Blazhko stars \citep{2011arXiv1106.4914B}.

\subsubsection{RR-PC stars}
RR-PC refers to period-changing stars. It's difficult to distinguish PC stars
from MC stars or BL stars since they all have closely spaced side frequencies.
\cite{2006A&A...454..257N} defined the PC stars as  that they have close
components which can not be eliminated within three prewhitening cycles or their
separation from the main pulsation component is $\leq \sim$1/T, where T is total
time span. The definition is followed in this work.  In our sample, we find that
no star has any frequency detected after four prewhitening loops, i.e. the
number of frequencies are not bigger than 4. The two stars which have a fourth
frequency component both have at least one frequency with the separation from
the main pulsation component $\leq \sim$1/T.  So they are classified as RR-PC
stars with no doubt. In addition, there are some RRLS which have fewer than four
frequency components but with some frequency whose separation from the main
component is $\leq \sim$1/T, the number of such stars is 18. Altogether, there
are 20 RR-PC stars, that is about 3 percent of the sample. Their variation
properties are listed in Table~\ref{tab7}. Our attempt to analyze the period variation is
hampered by the large interval between adjacent measurements which ranges from
0.003 d to 300 d with a mean value of about 3.6 d, that is to say, the interval
is several periods long.

\subsubsection{Other RRLS}

Eight (1.2\%) multi-period RRLS in our sample cannot be classified into the
above subtypes. They have more than one pulsation frequency, but their
frequencies are not closely spaced from the main frequency. Table~\ref{tab8} shows their
frequencies and other variation parameters.

Four of them have one period around 0.3~d and the other around 0.5~d, yielding a
period ratio around 0.6 which is the canonical period ratio between the SO and
FU modes \citep{1997ApJ...477..346B}. So we suspect that although we can't find
single SO mode pulsating RRLS, but they can co-exist with the FU mode. The four
RR02 candidates are shown by red triangles in Fig.~\ref{fig:3} where P$_{\rm
1st}$ and P$_{\rm 2nd}$ refer to the primary and secondary period respectively.
As same as the RR01 stars, the interaction between the two modes have led their
amplitude and period to be on the short/long end of the FU/SO mode, which bring
them together in the period-amplitude diagram (Fig.~\ref{fig:3}). Moreover, Star
228 seems to have a long modulation period for its FU mode from the fact that a
third frequency is found to be closely spaced to its FU frequency.

\begin{figure}
   \includegraphics[width=\textwidth]{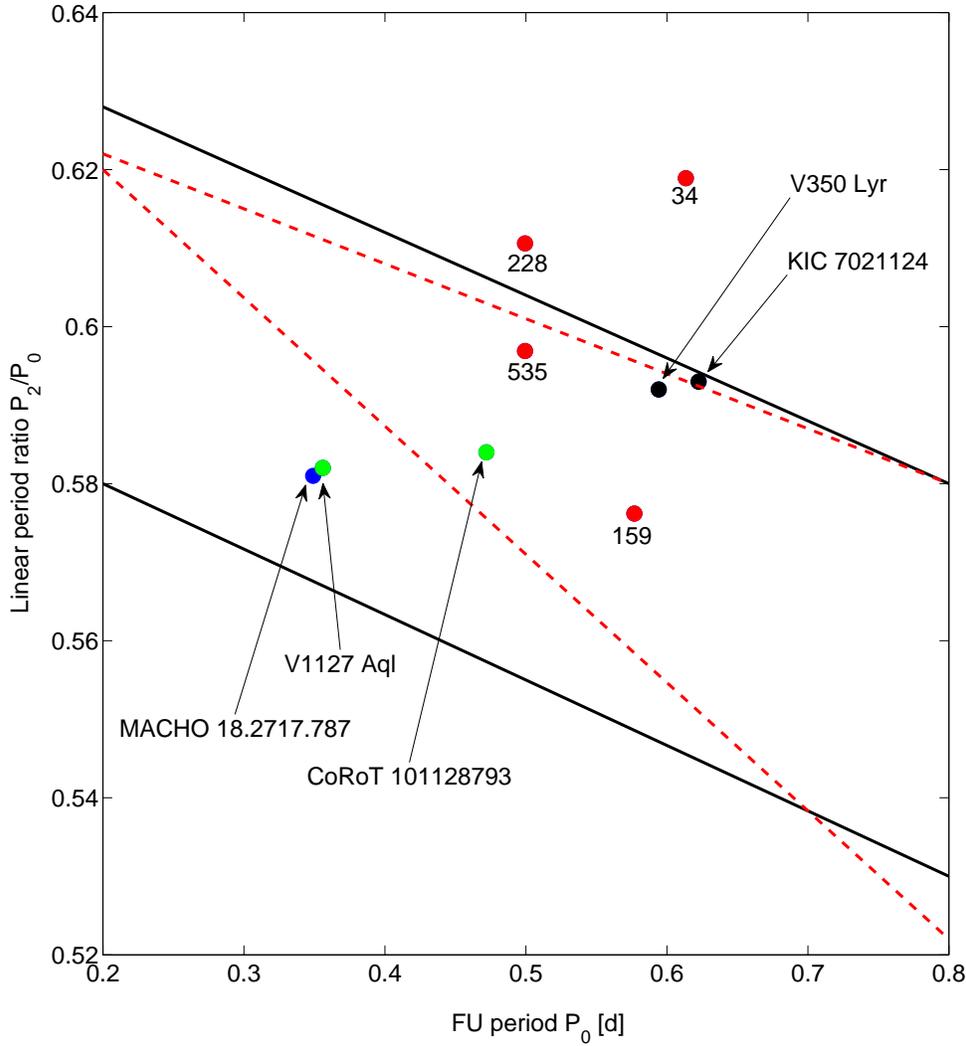}
\caption{The RR02 candidate stars in the Petersen diagram. The black solid lines
delineate the rough range of RR02 stars by models from
\citet{2010A&A...520A.108P}, and the red dash line from
\citet{2011arXiv1106.6120N}.  The blue, green, black and red dots show the
candidate RR02 stars from MACHO, CoRoT, Kepler datasets and our results from
OGLE III datasets, with the name or ID number labeled (see Section 6.4 for
details).}
\label{fig:11}
\end{figure}

Only four stars were clearly claimed to be RR02 stars in the recent two years.
All of them were discovered in space mission. They are V350 Lyr
\citep{2010MNRAS.409.1585B} and KIC 7021124 \citep{2011arXiv1106.6120N} from the
\textit{Kepler} mission, CoRoT 101128793 \citep{2010A&A...520A.108P} and V1127
Aql \citep{2010A&A...510A..39C} from the \emph{CoRoT} mission. We found that
star MACHO 18.2717.787 unidentified by \cite{2006A&A...454..257N} from MACHO
dataset could also be such double-mode RR02 star with the period ratio of
0.5810. \cite{2010A&A...520A.108P} computed a grid of linear RRLS models in a
large stellar parameter space which delineated a rough range of the RR02 stars
in the Petersen diagram. A similar work was presented by
\cite{2011arXiv1106.6120N} who used the Warsaw pulsation hydrocode including
turbulent convection. In Fig.~\ref{fig:11}, the range of RR02 in the Petersen
diagram defined by the two models are delimited by solid and dash lines
respectively. It can be seen that the two models agree with each other generally
but also disagree in particular at the short fundamental periods.  In this
diagram, Star MACHO 18.2717.787 denoted by a blue dot is definitely inside the
model range. The four RR02 candidates from our sample are shown by red dots with
other five such stars by dots in other colors. Star 159 and 535 are undoubtedly
inside the range by both models. Star 228 is just outside the upper border of
the models, but can not be excluded since the models surely have some
uncertainty. The only star which apparently deviates from the models is Star 34
although it is not too far. Puzzlingly, all the nine stars take a trend that the
period ratio increases with FU period, which is opposite to the models.
Interestingly, such discrepancy between observation and model occurs exactly the
same in RR01, the other double-mode stars (see Fig.5 of
\citealt{2000ApJ...542..257A}).

\begin{table}
\caption{Parameters of light variation for miscellaneous RRLS.}
\centering 
 \begin{tabular}{crrrrrrrrrrrr}
\hline\hline                        
ID	&	$P_0$	&	 $P_1$	&	
$P_2$	&	$A_0$	&	$\bar I$	&	
$P_{\rm short}/P_{\rm long}$	&	$A_{\rm shortP}/A_{\rm longP}$	 
&	Type	\\
\hline  
21	&		0.601141 	&	0.997709 	&	
&		0.466 	&	19.032 	 &	 0.6025		&	3.899
&	RR0-D1	 \\
34	&		0.379702 	&	0.613512 	&	
&		0.109 	&	18.667 	 &	 0.6189	 	&	1.846
&	RR02	 \\
159	&		0.576922 	&	0.332432 	&	
&		0.536 	&	18.733 	 &	 0.5762	 	&	0.185
&	RR02	 \\
181	&		0.564876 	&	0.815054 	&	
&		0.409 	&	18.191 	 &	 0.6931	 	&	4.964
&	?	 \\
228	&		0.304926 	&	0.499391 	&	0.499175
	&		0.231 	&	 18.599 	&	0.6106	
&	1.712	&	 RR-02	 \\
248	&		0.599241 	&	0.464902 	&	
&		0.413 	&	18.806 	 &	 0.7758	 	&	0.147
&	?	 \\
452	&		0.529150 	&	0.997639 	&	
&		0.445 	&	18.483 	 &	 0.5304	 	&	2.092
&	RR0-D1	 \\
535	&		0.298121 	&	0.499435 	&	
&		0.248 	&	18.655 	 &	 0.5969	 	&	2.879
&	RR02	 \\
\hline  
\end{tabular}
\label{tab8}
\end{table}

Two stars have a secondary period around 1 day. This frequency is not taken as the alias because the phased light-curves at this
frequency show reliable periodicity.  We hereby denote them as RR0-D1, following
\cite{2000ApJ...542..257A}. The other three stars can not be classified into
any of the classes described above. We just mark them by "?" in Table~\ref{tab8}.

\section{Discussion and summary}

The incidence rates of each subclass are shown in Table~\ref{tab9}. The majority, 85\%,
is single-mode pulsators. The RRLS exhibiting the Blazhko effect (sum of RR-BL1
and RR-BL2) is the second most numerous group. With 52 RR-BL stars, they consist
7.9\% of the sample. The incidence rates of Blazhko stars are compared with
previous results in Table~\ref{tab10}. As our identification of a frequency is quite
strict, the percentages in our sample should be taken as the lower limit of the
Blazhko incidence rate.

For RRLS in LMC, the Blazhko variables (sum of RR-BL2 and RR-BL1 stars) occur
less frequently in RR0 (7.5\%) than in RR1 stars (9.1\%). For RR1 stars, we can
see an increasing trend of the incidence rates from 2.0\% and 7.5\% in previous
work to 9.1\% in present work, this can be explained by the longer time span and
more precise data. But this trend does not show in RR0 stars although the data
has been improved as for RR1 stars. To further analyze the reason, it is found
that the incidence rate of RR-BL1 is comparable to previous work, the rate is
6.7\% for RR0-BL1 and 5.1\% for RR1-BL1, compared to 6.5\%
\citep{2003ApJ...598..597A} and 3.5\% \citep{2006A&A...454..257N}. But in
regarding to the RR-BL2 stars, the incidence rate of RR1-BL2 is comparable to
the work before, with 5.1\% to 3.5\%. Meanwhile the incidence rate of RR0-BL2 is
abnormally low, with only 0.8\% compared to 5.4\% in
\cite{2003ApJ...598..597A}. As mentioned in previous section, the reason may
lie on our very strict criteria to accept a frequency which can move a BL2 star
into a BL1 star in the case of high asymmetry of the amplitude in the two side
frequencies. This explanation finds support in the fact that the RR0-BL2 stars
have the amplitude ratio $A_{+}/A_{-}$ not far from unity.  Another possible
reason is that, among RR0-BL1 stars, 75\% (24 out of 32) have $f_+$, and all the
four RR0-BL2 stars have $A_+ >A_-$.  In the work of \cite{2002ASPC..259..396K},
80\% of RR0-BL1 stars were found to have $A_+$ component. We suspect that for
RR0-BL stars, there maybe some unknown effect to make $A_+$ much larger than
$A_-$, which made a lot of $A_-$ components missing. This effect does not appear
in RR1 stars, for example, \cite{2000ApJ...542..257A} found  37\% RR1-BL1 stars
have $A_-$ components, and in our sample only 43\% RR1-BL2 stars have $A_+ >
A_-$. Based on Eq.(45) from \cite{2011arXiv1106.4914B}, most of the RR0 stars have $\pi < \phi _m < 2 \pi$; while for those
RR1 stars, $\phi _m$ is evenly distributed between zero and $2 \pi$. 

People used to believe that the incident rates of the Blazhko variables are
lower in LMC than in the Galaxy. But this only suits for RR0 stars, may not be
true for RR1, as the work of \cite{2006A&A...454..257N} already suggested. From
our work, with the long time span of observation, the Blazhko incidence rate for
RR1 stars is larger in LMC than in the Galaxy bulge, as the rates are 5.1\% and
4.0\% for RR1-BL1 and RR1-BL2 respectively in our LMC sample in comparison with
the 3.1\% and 1.5\% from \cite{2003A&A...398..213M} or 2.9\% and 3.9\% from
\cite{2003AcA....53..307M} for the bulge RRLS sample. On the other hand, both
the improvement of the observational precision and the extension of
observational span increase the possibility to detect the Blazhko effect. Based
on the data from the \emph{Kepler} mission \cite{2010ApJ...713L.198K} and  the
Konkoly Blazhko  Survey  \citep{2009MNRAS.400.1006J}, the incidence rate can
exceed forty percent, but they are small samples with no more than 30 objects. Thus such comparison
may not be conclusive as the observation and the analysis techniques are not
uniform, and the samples are very different. 
According to these observations
of LMC, SMC, bulge and $\omega$ Cen, there is no clear relation between the incidence rate
and metallicity. What causes the difference in the Blazhko incidence rate in
different environments is unclear, which could be part of the difficulty in
understanding the mechanism for the Blazhko effect.

\begin{table}
%
\caption{Statistical results of the RRLS classification.}
\centering  
\begin{tabular}{ccrrcr}
\hline\hline                        

type	&	Short description	&	number
&	percent	&	subtype 	&	number	\\
\hline  
RR-SG & Single-period & 556	 
& 84.9\%	&	RR0-SG	&	424	\\
	&		&		&		&	RR1-SG	&
132	\\ \hline  
RR01	&	FU/FO double mode 	&	15	&	2.3\%	&	
&		\\ \hline  
RR02	&	FU/SO double mode 	&	4	&	0.6\%	&	
&		\\ \hline  
{RR-BL}	&	{One close component}	&	
{52}	&	{7.9\%}	&	RR0-BL1	&
32	\\
	&		&		&		&	RR1-BL1	&
9	\\
	&	{Two symmetric components}	&	
&		&	RR0-BL2	 &	4	\\
	&		&		&		&	RR1-BL2	&
7	\\ \hline  
{RR-MC}	&	{Multiple close components}
&	 {4}	&	{0.6\%}	&	RR0-MC
&	1	\\
	&		&		&		&	RR1-MC	&
3	\\ \hline  
{RR-PC}	&	{Period changing}	&	
{20}	&	{3.1\%}	&	RR0-PC	&
11	\\
	&		&		&		&	RR1-PC	&
9	\\ \hline  
RR-D1	&	Second frequency at unity 	&	2	&	0.3\%
&	RR0-D1	&	2	 \\ \hline  
RR-?	&	Mysterious double mode 	&	2	&	0.3\%	&
RR0-others	&	2	 \\
\hline  
\end{tabular}
\label{tab9}
\end{table}

\begin{table}
\centering
\caption{The incidence rates of Blazhko effect in the sample compared to
previous results. The references in the table are: (a):
\cite{2003A&A...398..213M}, (b): \cite{2003ApJ...598..597A}, (c):
\cite{2003AcA....53..307M}, (d): \cite{2006ApJ...651..197C}, (e):
\cite{2008CoAst.157..345M}, (f): \cite{2000ApJ...542..257A}, (g):
\cite{2006A&A...454..257N}, \textbf{(h): this work}.}
\begin{tabular}{crrrrrr}
\hline\hline    
Ref.	&	Bulge(a)	&	Bulge(c)	&
Bulge(d)	&	 LMC(b)	&	 $\omega$ Cen (e)	 &
\textbf{LMC(h)}	\\
\hline  
RR0	&	150	&	1942	&	1888	&	6391	&
70	&		\textbf{478}	 \\
RR0-BL1	&	16.7\%	&	12.5\%	&	8.8\%	&	6.5\%	&
4.3\%	&		 \textbf{6.7\%}	 \\
RR0-BL2	&	6.0\%	&	7.4\%	&	14.9\%	&	5.4\%	&
18.6\%	&		 \textbf{0.8\%}	 \\
RR0-MC	&		&	4.4\%	&	3.5\%	&	0.3\%	&	
&	\textbf{0.2\%}	 \\
RR0-PC	&		&		&	6.3\%	&	2.9\%	&	
&	\textbf{2.3\%}	\\
\hline  \hline  
Ref.	&	Bulge(a)	&	Bulge(c)	&	LMC(f)	&
LMC(g)	&	$\omega$ Cen (e)	 &		\textbf{LMC(h)}	\\
\hline  
RR1	&	65	&	771	&	1327	&	1332	&
81	&		\textbf{177}	\\
RR1-BL1	&	3.1\%	&	2.9\%	&	1.8\%	&	3.5\%	&
21.0\%	& \textbf{5.1\%}	 \\
RR1-BL2	&	1.5\%	&	3.9\%	&	0.2\%	&	4.0\%	&
4.9\%	& \textbf{4.0\%} \\
RR1-MC	&		&	5.3\%	&	0.4\%	&	1.0\%	&	
&		 \textbf{1.7\%}	\\
RR1-PC	&		&		&	10.6\%	&	14.0\%	&	
&		\textbf{4.5\%}	 \\
\hline  
\end{tabular}
\label{tab10}
\end{table}

\begin{acknowledgements}
We sincerely thank the OGLE team for their continuing efforts and generosity in
sharing data. We also thank the referee for his very helpful suggestions. This work is supported by the NSFC grant No. 10973004.
\end{acknowledgements}

\bibliographystyle{raa}
\bibliography{rrlyr-resub}

\begin{appendix}

\section{Data of the result of the analysis.} 

\begin{table}
\caption{Parameters of light variation for the RR-SG stars. The full table is available in electronic form at the CDS.}
\centering
\begin{tabular}{crrrrrr} 
\hline\hline                        
ID & Period & $\Theta _{\rm PDM}$ &  $A_0$
& $\bar{I}$ & subtype  \\
\hline  
1	&	0.611821	&	0.1361	&	0.559	&	18.845
&	RR0-SG	\\
2	&	0.408541	&	0.1926	&	0.261	&	18.551
&	RR1-SG	\\
4	&	0.555120	&	0.0749	&	0.806	&	18.720
&	RR0-SG	\\
5	&	0.497251	&	0.0814	&	0.755	&	18.889
&	RR0-SG	\\
6	&	0.581477	&	0.1681	&	0.514	&	18.733
&	RR0-SG	\\
8	&	0.551528	&	0.0953	&	0.638	&	18.807
&	RR0-SG	\\
9	&	0.515681	&	0.1468	&	0.412	&	18.464
&	RR0-SG	\\
10	&	0.545564	&	0.1076	&	0.669	&	18.807
&	RR0-SG	\\
\hline  
\end{tabular}
\label{tab2}
\end{table}

\begin{table}
\caption{Parameters of light variation for the RR01 stars.}
\centering 
\begin{tabular}{crrrrrrrr} 
\hline\hline                        
ID  & type & $P_0$ & $\Theta _{0}$ &
$A_{0}$ & $\bar I$  & $P_1/P_0$ & $\Theta
_{1}$  & $A_{1}/A_{0}$ \\
\hline  
3	&	RR01	&	0.464248 	&	0.61 	&	0.169
&	18.816 	&	0.7434	 &	0.59 	&	1.279	\\
53	&	RR01	&	0.463814 	&	0.55 	&	0.158
&	18.768 	&	0.7435	 &	0.50 	&	1.492	\\
157	&	RR01	&	0.465272 	&	0.57 	&	0.169
&	18.890 	&	0.7431	 &	0.48 	&	1.628	\\
174	&	RR01	&	0.493743 	&	0.57 	&	0.178
&	18.782 	&	0.7446	 &	0.45 	&	1.525	\\
249	&	RR01	&	0.511593 	&	0.62 	&	0.121
&	18.664 	&	0.7456	 &	0.39 	&	2.026	\\
264	&	RR01	&	0.461786 	&	0.51 	&	0.235
&	18.886 	&	0.7429	 &	0.57 	&	1.306	\\
330	&	RR01	&	0.460572 	&	0.47 	&	0.247
&	18.850 	&	0.7427	 &	0.59 	&	1.073	\\
332	&	RR01	&	0.468276 	&	0.73 	&	0.106
&	18.845 	&	0.7431	 &	0.37 	&	2.613	\\
382	&	RR01	&	0.481604 	&	0.84 	&	0.128
&	18.692 	&	0.7441	 &	0.57 	&	1.938	\\
483	&	RR01	&	0.534026 	&	0.65 	&	0.098
&	18.482 	&	0.7465	 &	0.26 	&	2.608	\\
528	&	RR01	&	0.459523 	&	0.60 	&	0.234
&	18.611 	&	0.7422	 &	0.50 	&	0.810	\\
531	&	RR01	&	0.465609 	&	0.65 	&	0.146
&	18.669 	&	0.7430	 &	0.47 	&	1.055	\\
558	&	RR01	&	0.486225 	&	0.52 	&	0.164
&	18.773 	&	0.7440	 &	0.49 	&	1.452	\\
622	&	RR01	&	0.462159 	&	0.47 	&	0.213
&	18.874 	&	0.7431	 &	0.55 	&	1.363	\\
642	&	RR01	&	0.468788 	&	0.57 	&	0.295
&	18.824 	&	0.7427	 &	0.42 	&	0.839	\\
\hline  
\end{tabular}
\label{tab3}
\end{table}

\begin{table}
%
\caption{Parameters of light variation for the RR-BL1 stars.}
\centering 
 \begin{tabular}{crrrrrrr}
\hline\hline                        
ID	  & 	type	  & 	$f_0$	  & 	
$\Theta _0$	  & 	$A_0	$  & 	$\bar I$	  &
	 $\Delta f$	  & 	$A_1/A_0$ \\
\hline  
7	&	RR0	&	1.545554	&	0.50 	&	0.106
&	18.056 	&	0.010905	 &	0.382	\\
20	&	RR0	&	1.592310	&	0.15 	&	0.747
&	18.597 	&	0.020002	 &	0.186	\\
78	&	RR0	&	1.792749	&	0.27 	&	0.515
&	18.975 	&	0.018073	 &	0.212	\\
103	&	RR0	&	1.791820	&	0.42 	&	0.419
&	18.770 	&	-0.001242	 &	0.381	\\
125	&	RR0	&	1.816095	&	0.37 	&	0.423
&	18.809 	&	0.009155	 &	0.308	\\
191	&	RR0	&	1.639542	&	0.27 	&	0.443
&	18.598 	&	-0.000641	 &	0.232	\\
193	&	RR0	&	2.148450 	&	0.25 	&	0.563
&	18.751 	&	0.005300 	 &	0.204	\\
196	&	RR0	&	1.703644	&	0.17 	&	0.388
&	18.496 	&	0.001861	 &	0.204	\\
217	&	RR0	&	1.941551	&	0.33 	&	0.477
&	18.910 	&	0.025364	 &	0.283	\\
230	&	RR0	&	2.056198	&	0.24 	&	0.594
&	18.768 	&	0.022642	 &	0.188	\\
239	&	RR0	&	1.596092	&	0.45 	&	0.265
&	18.696 	&	0.017372	 &	0.313	\\
253	&	RR0	&	1.814577	&	0.13 	&	0.583
&	18.586 	&	-0.004017	 &	0.119	\\
261	&	RR0	&	1.971605	&	0.24 	&	0.530
&	18.886 	&	0.007993	 &	0.213	\\
275	&	RR0	&	1.883136	&	0.17 	&	0.485
&	18.627 	&	0.035635	 &	0.138	\\
280	&	RR0	&	2.079711	&	0.23 	&	0.664
&	19.010 	&	0.043438	 &	0.203	\\
283	&	RR0	&	2.627460	&	0.31 	&	0.491
&	19.285 	&	0.016204	 &	0.279	\\
311	&	RR0	&	1.875675 	&	0.16 	&	0.645
&	18.733 	&	0.030124 	 &	0.217	\\
313	&	RR0	&	1.722907 	&	0.37 	&	0.393
&	18.765 	&	0.010309 	 &	0.338	\\
325	&	RR0	&	1.849624 	&	0.27 	&	0.543
&	18.882 	&	0.010813 	 &	0.256	\\
347	&	RR0	&	1.830031	&	0.12 	&	0.641
&	18.692 	&	-0.003638	 &	0.119	\\
377	&	RR0	&	1.587357	&	0.31 	&	0.293
&	18.736 	&	0.030942	 &	0.222	\\
427	&	RR0	&	1.777755	&	0.25 	&	0.502
&	18.721 	&	0.022702	 &	0.284	\\
444	&	RR0	&	1.969073	&	0.18 	&	0.667
&	18.862 	&	0.002295	 &	0.206	\\
464	&	RR0	&	1.717309	&	0.28 	&	0.377
&	18.766 	&	-0.009179	 &	0.175	\\
469	&	RR0	&	1.829113	&	0.25 	&	0.592
&	18.883 	&	0.016978	 &	0.272	\\
517	&	RR0	&	1.662463	&	0.42 	&	0.227
&	18.537 	&	0.026333	 &	0.345	\\
579	&	RR0	&	1.560733	&	0.21 	&	0.493
&	18.602 	&	-0.023590	 &	0.244	\\
597	&	RR0	&	1.323251	&	0.37 	&	0.197
&	18.507 	&	0.011120	 &	0.254	\\
624	&	RR0	&	1.671056	&	0.27 	&	0.360
&	18.391 	&	-0.007175	 &	0.286	\\
639	&	RR0	&	1.720961	&	0.25 	&	0.354
&	18.651 	&	0.019355	 &	0.200	\\
645	&	RR0	&	2.189412	&	0.15 	&	0.619
&	18.762 	&	0.014528	 &	0.173	\\
652	&	RR0	&	1.885687	&	0.24 	&	0.422
&	18.703 	&	-0.009030	 &	0.172	\\
95	&	RR1	&	3.295831	&	0.37 	&	0.352
&	19.148 	&	0.000348	 &	0.324	\\
107	&	RR1	&	3.310811	&	0.64 	&	0.110
&	18.716 	&	0.072968	 &	0.516	\\
126	&	RR1	&	3.246747	&	0.78 	&	0.087
&	18.737 	&	-0.072875	 &	1.081	\\
164	&	RR1	&	3.718042	&	0.66 	&	0.154
&	19.057 	&	0.132355	 &	0.728	\\
198	&	RR1	&	3.130009	&	0.47 	&	0.198
&	18.747 	&	0.000336	 &	0.340	\\
317	&	RR1	&	3.331974	&	0.47 	&	0.206
&	18.784 	&	-0.128505	 &	0.376	\\
322	&	RR1	&	2.739613	&	0.63 	&	0.178
&	18.734 	&	0.001267	 &	0.424	\\
542	&	RR1	&	2.889081	&	0.75 	&	0.073
&	18.700 	&	0.156416	 &	0.891	\\
613	&	RR1	&	3.684208	&	0.63 	&	0.142
&	19.009 	&	-0.072704	 &	0.482	\\

\hline  
\end{tabular}
\label{tab4}
\end{table}

\begin{table}
\caption{Parameters of light variation for the RR-BL2 stars.}
\centering  
\begin{tabular}{c c c c c c c c c c c}
\hline\hline                        
ID  & type & $f_0$  & $\Theta _0$ &
$A_0$ & $\bar I$ & $\Delta f_{+}$ &
$\Delta f_{-}$ & $A_+/A_0$ & $A_-/A_0$ &
$A_+/A_-$ \\
\hline  
165	&	RR0	&	2.150372 	&	0.27 	&	0.567
&	18.753 	&	0.022871	 &	0.022862	&	0.290
&	0.182	&	1.593	\\
237	&	RR0	&	2.044692 	&	0.44 	&	0.186
&	18.462 	&	0.000405	 &	0.000335	&	0.362
&	0.477	&	0.759	\\
441	&	RR0	&	1.545543 	&	0.32 	&	0.283
&	18.699 	&	0.016791	 &	0.016839	&	0.260
&	0.214	&	1.215	\\
460	&	RR0	&	1.594151 	&	0.13 	&	0.559
&	18.434 	&	0.000376	 &	0.000409	&	0.154
&	0.131	&	1.176	\\
594	&	RR0	&	1.741086 	&	0.42 	&	0.367
&	18.641 	&	0.000306	 &	0.000422	&	0.373
&	0.233	&	1.601	\\
92	&	RR1	&	2.988331 	&	0.45 	&	0.189
&	18.534 	&	0.000337	 &	0.000352	&	0.279
&	0.312	&	0.894	\\
227	&	RR1	&	3.646038 	&	0.38 	&	0.265
&	19.032 	&	0.000706	 &	0.000708	&	0.269
&	0.272	&	0.989	\\
357	&	RR1	&	3.560084 	&	0.48 	&	0.280
&	19.003 	&	0.016841	 &	0.016816	&	0.432
&	0.355	&	1.217	\\
529	&	RR1	&	3.699046 	&	0.69 	&	0.158
&	18.977 	&	0.082334	 &	0.082303	&	0.571
&	0.574	&	0.995	\\
564	&	RR1	&	3.031503 	&	0.64 	&	0.185
&	18.845 	&	0.000433	 &	0.000692	&	0.597
&	0.450	&	1.327	\\
580	&	RR1	&	3.114803 	&	0.41 	&	0.224
&	18.773 	&	0.006638	 &	0.006642	&	0.292
&	0.256	&	1.141	\\
\hline  
\end{tabular}
\label{tab5}
\end{table}

\begin{table}

\caption{Parameters of light variation for the RR-MC stars.}
\centering  
\begin{tabular}{crrrrrrrrrrrr}
\hline\hline                        
ID	&	type	&	$\Theta _{0}$	 &
	$A_0$	&	$\bar I$	&	$f_0$
&	$\Delta f_1	$ &	$\Delta f_2	$ &
$\Delta f_3	 $	& $A_1/A_0$	& $A_2/A_0$
&	 $A_3/A_0$	 &	Notes	\\
\hline  
219	&	RR0	&	0.50 	&	0.319 	&	18.675 	&
1.903236 	&	-0.001661 	 &	-0.000565 	&	0.000533
	&	0.568	&	0.341	&	0.303	&	 a	\\
156	&	RR1	&	0.53 	&	0.224 	&	18.810 	&
2.645540 	&	0.000353 	 &	-0.001795 	&
-0.000964 	&	0.428	&	0.335	&	0.330	&	 b
\\
161	&	RR1	&	0.52 	&	0.222 	&	18.742 	&
2.836165 	&	-0.001872 	 &	-0.000471 	&
-0.001482 	&	0.558	&	0.452	&	0.498	&	 c
\\
215	&	RR1	&	0.75 	&	0.176 	&	18.933 	&
2.817171 	&	0.000403 	 &	-0.000312 	&
-0.000804 	&	0.772	&	0.468	&	0.480	&	 b
\\
\hline  
\end{tabular}
\label{tab6}
\end{table}

\begin{table}
%
\caption{Parameters of light variation for the RR-PC stars.}
\centering
 \begin{tabular}{crrrrrrr} 
\hline\hline                        
ID	  & 	type	  & 	$f_0$	  & 	
$\Theta _0$	  & 	$A_0	$  & 	$\bar I$	  &
	 $\Delta f_i$	  & 	$A_i/A_0$\\
\hline  
413	&	RR0	&	1.717188 	&	0.51	&	0.298
&	18.576	&	-0.000309 	 &	0.590	\\
442	&	RR0	&	1.791474 	&	0.39	&	0.310
&	18.775	&	0.000204 	 &	0.278	\\
550	&	RR0	&	1.864087 	&	0.32	&	0.460
&	18.719	&	-0.000301 	 &	0.272	\\
496	&	RR1	&	3.656314 	&	0.71	&	0.147
&	18.955	&	-0.000279 	 &	0.503	\\
651	&	RR1	&	3.309307 	&	0.47	&	0.268
&	18.693	&	0.000244 	 &	0.378	\\
247	&	RR0	&	2.093590 	&	0.39	&	0.483
&	18.999	&	0.000253 	 &	0.246	\\
	&		&		&		&		&	
&	0.000192 	&	0.316	\\
473	&	RR1	&	2.447215 	&	0.58	&	0.221
&	18.514	&	0.001115 	 &	0.459	\\
	&		&		&		&		&	
&	0.000148 	&	0.319	\\
581	&	RR0	&	1.674840 	&	0.38	&	0.388
&	18.730	&	0.000257 	 &	0.226	\\
	&		&		&		&		&	
&	0.000428 	&	0.228	\\
28	&	RR1	&	2.981115 	&	0.74	&	0.152
&	19.091	&	-0.000213 	 &	0.686	\\
	&		&		&		&		&	
&	-0.000516 	&	0.648	\\
67	&	RR1	&	2.451110 	&	0.45	&	0.253
&	18.554	&	0.000265 	 &	0.388	\\
	&		&		&		&		&	
&	0.000235 	&	0.296	\\
297	&	RR1	&	3.127205 	&	0.67	&	0.229
&	18.713	&	0.000274 	 &	0.516	\\
	&		&		&		&		&	
&	0.000127 	&	0.501	\\
586	&	RR1	&	2.726628 	&	0.47	&	0.210
&	18.815	&	0.000278 	 &	0.327	\\
	&		&		&		&		&	
&	0.000204 	&	0.328	\\
407	&	RR0	&	2.141042 	&	0.27	&	0.471
&	19.002	&	0.000609 	 &	0.276	\\
	&		&		&		&		&	
&	0.000192 	&	0.219	\\
312	&	RR0	&	2.117445 	&	0.56	&	0.384
&	18.902	&	-0.000226 	 &	0.684	\\
	&		&		&		&		&	
&	0.000189 	&	0.217	\\
	&		&		&		&		&	
&	0.004992 	&	0.268	\\
495	&	RR0	&	2.075435 	&	0.23	&	0.410
&	18.597	&	0.000258 	 &	0.173	\\
	&		&		&		&		&	
&	-0.000092 	&	0.172	\\
	&		&		&		&		&	
&	0.017395 	&	0.161	\\
501	&	RR0	&	1.716627 	&	0.42	&	0.355
&	18.582	&	0.000232 	 &	0.362	\\
	&		&		&		&		&	
&	0.020464 	&	0.272	\\
	&		&		&		&		&	
&	-0.000269 	&	0.204	\\
176	&	RR1	&	2.743459 	&	0.71	&	0.169
&	18.627	&	-0.000926 	 &	0.459	\\
	&		&		&		&		&	
&	-0.000240 	&	0.436	\\
	&		&		&		&		&	
&	-0.000642 	&	0.492	\\
519	&	RR1	&	2.662524 	&	0.40	&	0.249
&	18.678	&	-0.000535 	 &	0.311	\\
	&		&		&		&		&	
&	-0.000840 	&	0.262	\\
	&		&		&		&		&	
&	0.000234 	&	0.236	\\
435	&	RR0	&	1.639894 	&	0.61	&	0.283
&	18.684	&	0.000207 	 &	0.513	\\
	&		&		&		&		&	
&	0.000161 	&	0.633	\\
	&		&		&		&		&	
&	0.000090 	&	0.280	\\
	&		&		&		&		&	
&	0.000354 	&	0.336	\\
596	&	RR0	&	1.369461 	&	0.56	&	0.279
&	18.713	&	0.000316 	 &	0.637	\\
	&		&		&		&		&	
&	0.000064 	&	0.523	\\
	&		&		&		&		&	
&	0.000165 	&	0.404	\\
	&		&		&		&		&	
&	0.000180 	&	0.272	\\
\hline  
\end{tabular}
\label{tab7}
\end{table}

\end{appendix}
\label{lastpage}

\end{document}